\documentclass[journal,twoside,onecolumn,letterpaper]{IEEEtran}
\usepackage{ifpdf}
\usepackage{color}
\usepackage[latin1]{inputenc}
\usepackage{cite}
\usepackage{mathtools}
\usepackage{amssymb}
\usepackage{algorithm}
\usepackage{algpseudocode}
\usepackage[artemisia]{textgreek}
\usepackage{url}

\usepackage{subfigure}

\usepackage{graphicx} \DeclareGraphicsExtensions{.png}

\usepackage{tikz}
\usetikzlibrary{decorations.pathmorphing} % noisy shapes
\usetikzlibrary{fit}					% fitting shapes to coordinates
\usetikzlibrary{backgrounds}	% drawing the background after the foreground
\usetikzlibrary{shapes}
\usetikzlibrary{positioning}
\usetikzlibrary{shadows}
\usetikzlibrary{decorations.pathmorphing}
\usetikzlibrary{decorations.shapes}
\usepackage{multirow}
\usepackage{dcolumn}
\usepackage{pgfplots}
\pgfplotsset{compat=1.3}% <-- moves axis labels near ticklabels (respects tick label widths)

\newtheorem{theorem}{Theorem}
\newtheorem{lemma}{Lemma}

\newtheorem{corollary}{Corollary}

\begin{document}
\allowdisplaybreaks

%page numbers on top with chapter heading
\pagestyle{headings}
%\title{Poisson Point Process Functionals with Applications to Interference in Wireless Networks}
\title{Interference Functionals in Poisson Networks}
\author{Udo Schilcher, Stavros Toumpis, Martin Haenggi, Alessandro Crismani, G\"unther Brandner, Christian Bettstetter
\thanks{Udo Schilcher, G\"unther Brandner, and Christian Bettstetter are with the Institute of Networked and Embedded Systems, University of Klagenfurt, Klagenfurt, Austria. E-Mail: \texttt{udo.schilcher@uni-klu.ac.at}}
\thanks{Stavros Toumpis is with the Department of Informatics, Athens University of Economics and Business, Athens, Greece.}
\thanks{Martin Haenggi is with the Department of Electrical Engineering, University of Notre Dame, IN, USA and with the \'Ecole Polytechnique F\'ed\'erale de Lausanne (EPFL), Switzerland.}
\thanks{Alessandro Crismani was with the Institute of Networked and Embedded Systems, University of Klagenfurt, Klagenfurt, Austria, and is now with u-blox, Trieste, Italy.}

        %\texttt{alessandro.crismani@ieee.org, udo.schilcher@aau.at, guenther.brandner@aau.at,} \\
        %\texttt{toumpis@aueb.gr, christian.bettstetter@auu.at}
}

\markboth{}{}
%\pubid{0000--0000/00\$00.00~\copyright~2007 IEEE}
\thispagestyle{empty}

% formatting equations in one or two columns
\newboolean{twoc}
\ifCLASSOPTIONonecolumn
	\setboolean{twoc}{false}
\else
	\setboolean{twoc}{true}
\fi

\makeatletter
\newcommand{\vast}{\bBigg@{3.5}}
\newcommand{\Vast}{\bBigg@{4}}
\makeatother

\newcommand{\eqleft}{\ifthenelse{\boolean{twoc}}{\lefteqn}{}}
\newcommand{\eqnln}{\ifthenelse{\boolean{twoc}}{\\\nonumber&=&}{&=&}}

% notation
\newcommand{\etal}{{et al.\;}}
\newcommand{\eul}{{\rm e}}

\newcommand{\expect}{{\rm E}}
\newcommand{\n}{\mathbb{N}}
\newcommand{\nwith}{\mathbb{N}_0}
\newcommand{\z}{\mathbb{Z}}
\newcommand{\re}{\mathbb{R}}

\newcommand{\sinc}{\mathrm{sinc}}

\newcommand{\usum}{\sum^{\neq}}

\newcommand{\origin}{o}		% origin of coordinate system

\newcommand{\BesselI}{B}

\newcommand{\sir}{\mathrm{SIR}}

\newcommand{\sigmaalg}{{\cal A}}
\newcommand{\prob}{{\rm P}}
\newcommand{\pset}{N}
\newcommand{\psigmaalg}{{\cal N}}
\newcommand{\x}{X}
\newcommand{\xmeas}{{\cal X}}
\newcommand{\rvspace}{Y}
\newcommand{\rvsigmaalg}{{\cal Y}}
\newcommand{\meas}{\mu}
\newcommand{\measset}{{\cal S}}

\newcommand{\indic}{{\rm\bf 1}}
\newcommand{\dd}{{\rm d}}
\newcommand{\euler}{\mathrm{e}}

\newcommand{\ppp}{\Phi}
\newcommand{\pppel}{\phi}
\newcommand{\pppre}{\varphi}
\newcommand{\pppn}{{\cal N}}
\newcommand{\pppnk}{{{\cal N}_\ppppc}}
\newcommand{\pppp}{P}
\newcommand{\ppppc}{K}
\newcommand{\ppppcrange}{[\ppppc]}
\newcommand{\dens}{\lambda}
\newcommand{\densf}{\Lambda}

\newcommand{\src}{S}
\newcommand{\dst}{D}
\newcommand{\thr}{\theta}
\newcommand{\thri}{\thr_{\mathit{\scriptscriptstyle\src\dst}}}
\newcommand{\gain}{{\ell}}
\newcommand{\gains}{{\gain_\mathrm{s}(u)}}
\newcommand{\gainm}{{\gain_\mathrm{m}(u)}}
\newcommand{\gaine}{{\gain_\epsilon(u)}}
\newcommand{\gaind}{{\gain_\mathrm{d}(u)}}

\newcommand{\fgain}{h}
\newcommand{\plexp}{\alpha}
\newcommand{\falpha}{\delta}
\newcommand{\pleps}{\epsilon}
\newcommand{\txp}{\wp}
\newcommand{\interf}{I}

\newcommand{\gth}{\psi}
\newcommand{\rilam}{\psi}
\newcommand{\fadk}{k}
\newcommand{\nakagamim}{m}

\newcommand{\brad}{a}
\newcommand{\ball}{{b(\origin,\brad)}}
\newcommand{\dimensions}{n}
\newcommand{\fcount}{q}
\newcommand{\pppspace}{{\mathbb{R}^\dimensions}}
\newcommand{\pppspaced}{{\mathbb{R}^{\dimensions\dpwr}}}
\newcommand{\pwr}{k}
\newcommand{\f}{f}
\newcommand{\fcum}{f^{\otimes}}
\newcommand{\g}{g}
\newcommand{\h}{f}
\newcommand{\perm}{\pi}

%indices in many different equations
\newcommand{\indi}{i}
\newcommand{\indii}{j}
\newcommand{\indiii}{k}
\newcommand{\indiv}{r}
\newcommand{\indv}{s}
\newcommand{\indlen}{l}
\newcommand{\maxlens}{\min(\|\fexp\|_1,|\someset|)}
\newcommand{\maxlen}{\min(\|\fexp\|_1,|\ppp|)}
\newcommand{\maxlenk}{\min(\|\fexp\|_1,\ppppc)}
\newcommand{\maxlenhomo}{\|\fexp\|_1}

\newcommand{\pointi}{u}
\newcommand{\pointii}{v}
\newcommand{\pointiii}{w}
\newcommand{\coord}{x}
\newcommand{\natvec}{N}

\newcommand{\someset}{U}
\newcommand{\somesetii}{V}
\newcommand{\rndvar}{X}
\newcommand{\rndvardom}{{\cal X}}
\newcommand{\mat}{M}
\newcommand{\matset}{{\cal M}_l^p}
\newcommand{\matseti}{{\cal M}_l^{(\indi)}}
\newcommand{\matsetii}{{\cal M}_l^{(\indi,\indii)}}
\newcommand{\melem}{m}
\newcommand{\col}{c}
\newcommand{\row}{r}

\newcommand{\magic}{{\cal C}}

\newcommand{\fexp}{p}
\newcommand{\fexpc}{p^c}
\newcommand{\pdf}{f}

\newcommand{\slot}{t}
\newcommand{\cperm}{d}

\newcommand{\dpwr}{d}

\newcommand{\sevent}{A}

% constants
\newcommand{\const}{c}

% indicator functions
\newcommand{\indicator}{{\rm\bf 1}}
\newcommand{\indici}{\indicator\big(\pointi\in\ppp_\indi)}
\newcommand{\indicii}{\indicator\big(\pointii\in\ppp_\indii)}
\newcommand{\indicxii}{\indicator\big(\coord\in\ppp_\indii)}
\newcommand{\indicxiii}{\indicator\big(\coord\in\ppp_\indiii)}
\newcommand{\indiciv}{\indicator\big(\pointi\in\ppp_\indiv)}
\newcommand{\indicv}{\indicator\big(\pointii\in\ppp_\indv)}
\newcommand{\indicwiv}{\indicator\big(\pointiii\in\ppp_\indiv)}
\newcommand{\indicwv}{\indicator\big(\pointiii\in\ppp_\indv)}

\newcommand{\indim}{{\indicator(\melem_{\indi\indii}=0)}}
\newcommand{\indimi}{{\indicator(\melem_{\indiii 1}=0)}}
\newcommand{\indimii}{{\indicator(\melem_{\indiii 2}=0)}}

\newcommand{\isom}{F}

\newcommand{\scut}{\kappa}

\newcommand{\dist}{d}

\sloppy

\maketitle
\thispagestyle{empty}

\begin{abstract}

We propose and prove a theorem that allows the calculation of a class of functionals on Poisson point processes that have the form of expected values of sum-products of functions.
In proving the theorem, we present a variant of the Campbell-Mecke theorem from stochastic geometry.
We proceed to apply our result in the calculation of expected values involving interference in wireless Poisson networks. Based on this, we derive outage probabilities for transmissions in a Poisson network with Nakagami fading. Our results extend the stochastic geometry toolbox used for the mathematical analysis of interference-limited wireless networks. 

\end{abstract}

\section{Introduction and Contributions}

%\PARstart{M}{ultiple}-access 
\PARstart{I}{nterference} in wireless networks occurs if the communication from a transmitter to a receiver is disturbed by additional nodes transmitting in the vicinity of the receiver on the same frequency band and at the same time.
Interference occurs even if code division multiple access (CDMA) is used, in which case multiple users are not separated in time or space but though the use of spreading codes; in this case, as well, interference powers, albeit reduced by the spreading, add up at the receiver.
Interference can be mitigated or even exploited~\cite{Katti07:interf-exp,kellogg14:backscatter} in networks with central entities using scheduling and signal processing techniques, such as multiuser detection~\cite{honig95:MU} and interference cancellation~\cite{andrews05:ic}. In non-centralized systems, however, when no dedicated control entities can regulate the access to the shared wireless medium, interference remains a performance-limiting factor, partly because it is subject to considerable uncertainty~\cite{haenggi09:interference}. 

For these reasons, the stochastic modeling of interference in wireless networks\,---\,in particular its dynamic behavior over time and space \cite{ganti09:interf-correl,schilcher12:interfcor,tanbourgi14:mrc}, which we will refer to in this work as the \emph{interference dynamics}\,---\, has recently attracted the interest of the research community. Temporal and spatial dependencies introduced by interference cause the events that different transmissions are correctly decoded to be dependent, which in turn influences system performance. It leads to reduced diversity~\cite{haenggi13:div-poly} and degraded performance of many communication techniques, such as cooperative relaying, multiple-input multiple-output (MIMO), and medium access protocols~\cite{schilcher2013coop,tanbourgi14:mrc,tanbourgi14:nakagami}. 

When modeling interference and its dynamics in wireless networks, researchers often use tools from stochastic geometry~\cite{net:Haenggi09jsac}. These tools include the Campbell-Mecke theorem, Campbell's theorem, and expressions for the probability generating functional (pgfl) (see~\cite{haenggi13:book, stoyan95, cressie93}) applied to Poisson point processes (PPPs). A comprehensive overview on applying stochastic geometry to the analysis of wireless networks can be found in books by Baccelli and B{\l}aszczyszyn~\cite{baccelli09:vol1,baccelli09:vol2} and by Haenggi \cite{haenggi13:book}.

%\comment{TODO: Write sentence about Th. 2 here AND in abstract.}

The article at hand extends the tools of stochastic geometry by calculating general sum-product functionals for PPPs.
While proving our results, we present a variant of the Campbell-Mecke theorem applied to PPPs. 
%These theorems enable us to calculate certain expected values over PPPs. 
Furthermore, we apply our results to calculate expected values that occur in the analysis of the interference in wireless networks. Notably, we derive link outage probabilities in a Poisson network with Nakagami fading~\cite{nakagami60:fading} caused by multipath propagation.

\subsection{Related Work}
In networks, in which the nodes' locations are modeled via a Poisson point process and that employ ALOHA, the interferers' locations form a Poisson point process. Therefore, we call such networks Poisson networks. This class of networks was the first in which the correlation of interference levels at different times and locations was analytically studied.% The paragraph summarizes the most important results:
Notably, mathematical expressions for interference dynamics are presented in \cite{ganti09:interf-correl,haenggi09:outage,schilcher12:interfcor,schilcher13:scc}. Analytical studies of cooperative diversity under correlated interference are performed in \cite{schilcher2013coop,crismani14:icc,tanbourgi14:mrc,crismani14:tvt,tanbourgi14:nakagami} using different assumptions concerning diversity combining (selection combining and maximum ratio combining) and small-scale fading (Rayleigh and Nakagami fading).% All these articles assume a \emph{Poisson network} model, i.e., the nodes are placed according to a PPP and no carrier sensing is employed for medium access. This set of assumptions permits the derivation of analytic expressions for outage probabilities. 
In Poisson networks, many different research works study the impact of interference dynamics on network performance.
The article~\cite{weber07:it} investigates the effects of the channel model and scheduling on the SIR, the outage probability, and the transmission capacity. From a more theoretical perspective, the diversity gain of retransmissions under correlated interference is analyzed in~\cite{haenggi13:div-poly} by means of diversity polynomials. Results show that the diversity gain is equal to unity even for lightly correlated interference despite independently fading channels in each transmission. The impact of node mobility on interference dynamics is investigated in~\cite{gong14:tmc}. The authors conclude that mobility reduces the temporal correlation of interference, but also has other implications on its statistics (e.g., it changes the expected value of interference).

Similar results can also be derived for cellular networks, if we assume Poisson distributed base stations. 
In this case, as well, interference is correlated across time and space, and its dynamics should be carefully taken into consideration when designing a system.
For example, in~\cite{haenggi14:cellular} it is shown how inter-cell interference coordination and intra-cell diversity impact the performance of a cellular network. In particular, it is shown that, depending on the SIR regime under consideration, one or the other of these techniques should be selected. Further, in~\cite{heanggi14:multipoint} coordinated multipoint transmissions in cellular networks are analyzed; by employing the coverage probability as a performance metric, the authors show that multipoint transmissions are more beneficial for the worst-case user than for the average user.

There is currently not as much theoretical work available on interference dynamics for Poisson networks employing carrier-sense multiple access (CSMA). When modeling CSMA, a minimum distance between sending nodes is introduced, which is typically modeled by hard-core processes. 

Mat\'ern's model is based on a dependent thinning process of a PPP, where each point is marked with a random number and the point with the highest number within a certain range is sustained; the others are removed. It is applied for modeling CSMA networks in many different publications: The authors of \cite{baccelli09:vol2} derive some theoretical results on this modeling assumption, although a comprehensive analysis of the interference dynamics is still subject to future work. The mean interference in CSMA networks is discussed in~\cite{haenggi11:mean-interf}. 
An analysis of coverage and throughput per user in IEEE 802.11 networks based on Mat\'ern's model is presented in~\cite{nguyen07:csma}; the results of this analysis are also used to solve some optimization problems. In~\cite{alfano11:csma} an analysis of dense CSMA networks is presented; the authors employ performance metrics such as average throughput to show that different spatial models lead to a significant change in network performance.

Approaches not based on Mat\'ern's model have also been considered. For example, a modified hard-core point process is proposed in~\cite{elsawy12:csma} to model the transmitters in a CSMA network; the authors derive closed-form solutions that approximate mean and variance of the interference; a simulation shows the accuracy of their approximations.
Also, in~\cite{busson09:ssi} the authors discuss how to model the spatial distribution of transmitting nodes showing that simple sequential inhibition processes are more suitable for modeling CSMA networks than Mat\'ern's model.

%The impact of correlated interference on 
The question as to how to design protocols that take advantage of the knowledge about interference correlation is still a rather unsolved problem. For example, in~\cite{zhong14:twc} its impact on MAC protocol design is discussed.
We therefore work toward a better understanding of the impact of correlated interference by presenting very general results on interference functionals.

In the article at hand we present mathematical tools that allow researchers to further generalize the modeling assumptions when studying interference dynamics. This allows to gain insights on interference dynamics in realistic scenarios.

\subsection{Summary of Contributions}
The main contributions of this article are as follows. Firstly, in Section~\ref{sec:theory} we provide and prove a theorem for calculating a functional that has the form of the expected value of a combined sum and product of functions over a PPP. 
In particular, let us consider a PPP $\ppp$ on $\pppspace$ with a locally finite and diffuse intensity measure $\densf$ and the $\fcount+1$ non-negative measurable functions $\f_\indi,\g:\pppspace\times\rndvardom\rightarrow\re$ with $\rndvardom\subseteq\re$, $i\in[\fcount]=\{1,\dots,\fcount\}$, and $\g(\coord,\chi)\leq 1$ for all $\coord\in\pppspace$ and $\chi\in\rndvardom$. Let the exponents $\fexp_\indi$ for $\indi\in[\fcount]$ ($\fcount\in\n$) be non-negative integers. We assume $0\in\n$ throughout the article. Finally, let $\rndvar=(\rndvar_\pointi)$ for all $\pointi\in\ppp$ be $\rndvardom$-valued i.i.d.~random marks to the points in $\ppp$. These marks can be used to model, e.g., the effects of multi-path propagation in an wireless network. Throughout this article, the notation $\expect_X$ indicates that the expected value is calculated with respect to $X$. Our contribution is to calculate the functional
\begin{equation}\label{eq:goal}
\expect_{\ppp,\rndvar}\left[\prod_{\indi=1}^\fcount\left(\sum_{\pointi\in\ppp}\f_\indi(\pointi,\rndvar_\pointi)\right)^{\fexp_\indi}\prod_{\pointi\in\ppp}\g(\pointi,\rndvar_\pointi)\right] \:.
\end{equation}
In the following, we call functionals of this form \emph{sum-product functionals}.
The expression we arrive at is given in \eqref{eq:th6} of Theorem~\ref{th6}. As part of the proof, we also present a variant of the Campbell-Mecke theorem applied to PPPs in Lemma~\ref{th:CMgen}.

Secondly, in Section~\ref{sec:expect} we apply this result to derive expressions for certain functionals related to interference in wireless networks with slotted ALOHA. In particular, we consider a setting where the interference power at the origin $\origin$ at time slot $\indi$ is $\interf_\indi=\sum_{\pointi\in\ppp}\fgain_\indi(\pointi)\,\gain(\pointi)\indici$ with $\fgain_\indi(\pointi)$ being the fading coefficient for node $\pointi$ at time slot $\indi$ (assumed to be temporally and spatially i.i.d.), $\gain(\pointi)$ being the path gain of node $\pointi$, and $\indici$ being the indicator function with $\ppp_\indi\subseteq\ppp$ denoting the nodes transmitting at time slot $\indi$.
 We are able to calculate 
\begin{equation}\label{eq:expectmain:intro}
\expect_{\ppp,\fgain,\indicator}\left[\prod_{\indi=1}^\fcount\interf_\indi^{\fexp_\indi}\,\exp\big(-\const \interf_\indi\big)\right]
\end{equation}
for some constant $c\in\re$ and $\fexp=(\fexp_1,\dots,\fexp_\fcount)\in\n^\fcount$. We call functionals of this form \emph{interference functionals}.
The result is given in~\eqref{eq:expectmain} of Theorem~\ref{th:expectmain}.
We then highlight some special cases particularly useful to wireless communications by employing commonly used models for path loss and small-scale fading into the general result of~\eqref{eq:expectmain:intro}. For example, for a stationary PPP, the singular path loss model $\gains=\|\pointi\|_2^{-\plexp}$ with path loss exponent $\plexp$, Rayleigh fading, and slotted ALOHA, we obtain
\begin{equation}\label{eq:expect:ex:intro}
\expect_{\ppp,\fgain,\indicator}[I \exp(-I)]=\frac{\falpha^2\dens\pi^2\exp\left(-\frac{\falpha\dens\pi^2}{\sin\left(\falpha\pi\right)}\right)}{\sin\left(\falpha\pi\right)}\:
\end{equation}
with $\falpha=\frac{2}{\plexp}$.
Other examples can be obtained by substituting the expressions presented in Table~\ref{tab:fadingexpect} into~\eqref{eq:i_expi_gen}.

Finally, in Section~\ref{sec:nakagami} we show how these results can be used in the performance analysis of wireless networks. For example, we derive the probability of a successful reception in a Poisson network with Nakagami fading, where the reception is successful iff the signal-to-interference ratio ($\sir$) is above a certain threshold $\thr$. The result is given in \eqref{eq:nakagamii}. We also derive the joint probability for the successful reception of two transmissions at the same receiver in different time slots.

\section{Sum-Product Functionals on PPPs}\label{sec:theory}

\subsection{Theorems from Stochastic Geometry}

Let $\ppp$ denote a PPP on $\pppspace$ with a locally finite and diffuse intensity measure $\densf$.
Stochastic geometry provides a set of helpful tools to calculate certain expected values involving $\ppp$.
One well-known tool is Campbell's theorem (see~\cite{haenggi13:book}, Section~4.5), which states that
\begin{equation}\label{eq:campbell}
\expect_\ppp\Bigg[\sum_{\pointi\in\ppp}\f(\pointi)\Bigg] = \int_{\pppspace}\f(\coord)\,\densf(\dd \coord)
\end{equation}
for any non-negative measurable function $\f$ on $\pppspace$.
Another tool is the following theorem on probability generating functionals (pgfls) of PPPs (see~\cite{haenggi13:book}, Section~4.6):
\begin{equation}\label{eq:pgfl}
\expect_\ppp\Bigg[\prod_{\pointi\in\ppp}\g(\pointi)\Bigg] = \exp\bigg(-\int_{\pppspace}\big(1-\g(\coord)\big)\,\densf(\dd \coord)\bigg)
\end{equation}
for any measurable function $\g$ on $\pppspace$ with $0\leq \g(\coord)\leq 1$ for all $\coord$ such that the integral in~\eqref{eq:pgfl} is finite.
A third tool is the following form of the Campbell-Mecke theorem (see \cite{haenggi13:book}, Section~8.4), which is a combination of~\eqref{eq:campbell} and~\eqref{eq:pgfl} and states
\begin{eqnarray}\label{eq:mecke}
\expect_\ppp\left[\sum_{\pointi\in\ppp}\f(\pointi)\prod_{\pointii\in\ppp}\g(\pointii)\right]=\exp\left(-\int_\pppspace \big(1-\g(\coord)\big)\,\densf\big(\dd\coord\big)\right)
\int_\pppspace \f(\coord)\g(\coord)\,\densf\big(\dd\coord\big)\:.
\end{eqnarray}
% \begin{equation}\label{eq:mecke}
% \expect\left[\sum_{\pointi\in\ppp}\f(\pointi)\prod_{\pointii\in\ppp}\g(\pointii)\right]=\exp\left(-\int_\pppspace \big(1-\g(\coord)\big)\,\densf\big(\dd\coord\big)\right)
% \int_\pppspace \f(\coord)\g(\coord)\,\densf\big(\dd\coord\big)\:.
% \end{equation}

These theorems are very helpful in the analysis of wireless networks and many other systems. None of them, however, can be applied to calculate an expected value of the form given in~\eqref{eq:goal}.
%\begin{equation}\label{eq:goal}
%\expect_{\ppp,\rndvar}\left[\prod_{\indi=1}^\fcount\left(\sum_{\pointi\in\ppp}\f_\indi(\pointi,\rndvar_\pointi)\right)^{\fexp_\indi}\prod_{\pointi\in\ppp}\g(\pointi,\rndvar_\pointi)\right] \:.
%\end{equation}
Such expected values are required in the analysis of wireless networks with interference, e.g., when calculating the outage probabilities under Nakagami fading. We therefore provide an expression for \eqref{eq:goal} in this section. 

\subsection{Higher-order Campbell-Mecke Theorem for PPPs}

In the following we prove two variants of the higher-order Campbell-Mecke theorem~\cite{haenggi13:book}  for PPPs (which is a special case of Corollary~9.2.3 in~\cite{baccelli09:vol1}).
\begin{lemma}[Higher-order Campbell-Mecke Theorem for PPPs]\label{th:CMgen}\it
Let $\ppp$ denote a PPP with a locally finite and diffuse intensity measure $\densf$. Further, let $\h:\pppspaced\times\pppn\rightarrow\re^+$ denote a measurable function, where $\pppn$ is the space of counting measures. Then we have
\begin{eqnarray}
\expect_\ppp\left[\usum_{\pointi\in\ppp^\dpwr}\h(\pointi,\ppp)\right]=\int_\pppspace\!\dots\int_\pppspace\int_\pppn \h(\coord,\pppre) \pppp_{\{\coord_1,\dots,\coord_\dpwr\}}(\dd \pppre)\densf(\dd \coord_1)\cdots\densf(\dd \coord_\dpwr)\:,
\end{eqnarray}
where the symbol $\neq$ on top of the sum denotes summing only over $\pointi=(\pointi_1,\dots,\pointi_\dpwr)$ with $\pointi_\indi\neq\pointi_\indii$ for all $\indi\neq\indii$. Further, $\coord=(\coord_1,\ldots,\coord_\dpwr)$ and $\pppp_{\{\coord_1,\dots,\coord_\dpwr\}}$ denotes the Palm distribution of $\ppp$ with respect to the points $\coord_1, \dots,\coord_\dpwr$. 
\end{lemma}
\begin{IEEEproof}
We get the result by iteratively applying the Campbell-Mecke theorem~\cite{haenggi13:book}:
\begin{eqnarray}
\expect_\ppp\left[\usum_{\pointi\in\ppp^\dpwr}\h(\pointi,\ppp)\right]
&\stackrel{(a)}{=}&\expect_\ppp\left[\sum_{\pointi_1\in\ppp}\sum_{\pointi_2\in\ppp\backslash\{\pointi_1\}}\cdots\sum_{\pointi_\dpwr\in\ppp\backslash\{\pointi_1,\dots,\pointi_{\dpwr-1}\}}\h\big((\pointi_1,\dots,\pointi_\dpwr),\ppp\big)\right]\\\nonumber
&\stackrel{(b)}{=}&\int_\pppn\sum_{\pointi_1\in\ppp}\sum_{\pointi_2\in\ppp\backslash\{\pointi_1\}}\cdots\sum_{\pointi_\dpwr\in\ppp\backslash\{\pointi_1,\dots,\pointi_{\dpwr-1}\}}\h\big((\pointi_1,\dots,\pointi_\dpwr),\pppre\big)\pppp(\dd \pppre)\\\nonumber
&\stackrel{(c)}{=}&\int_\pppspace\int_\pppn\sum_{\pointi_2\in\pppre}\cdots\sum_{\pointi_\dpwr\in\pppre}\h\big((\coord_1,\pointi_2,\dots,\pointi_\dpwr),\pppre\big)\pppp_{\{\coord_1\}}(\dd \pppre)\densf(\dd \coord_1)\\\nonumber
&\stackrel{(d)}{=}&\dots=\int_\pppspace\cdots\int_\pppspace\int_\pppn\h\big((\coord_1,\dots,\coord_\dpwr),\pppre\big)\pppp_{\{\coord_1,\dots,\coord_\dpwr\}}(\dd \pppre)\densf(\dd \coord_1)\cdots\densf(\dd \coord_\dpwr)\:,
\end{eqnarray}
where $(a)$ holds due to the definition of $\usum$; in $(b)$ we substitute the definition of the expected value over $\ppp$; and in $(c)$ and $(d)$ we apply the Campbell-Mecke theorem.
\end{IEEEproof}

\begin{corollary}\label{cor:CMgen}\it
With the same assumptions as in Lemma~\ref{th:CMgen} and $\h:\pppspace\times\pppn\rightarrow\re^+$ we have
\begin{equation}
\expect_\ppp\left[\sum_{\someset\subseteq\ppp\atop|\someset|=\dpwr}\prod_{\pointi\in\someset}\h(\pointi,\ppp)\right]=\frac{1}{\dpwr!}\,\int_\pppspace\!\dots\int_\pppspace\int_\pppn \prod_{\indi=1}^\dpwr \h(\coord_\indi,\pppre) \pppp_{\{\coord_1,\dots,\coord_\dpwr\}}(\dd \pppre)\densf(\dd \coord_1)\cdots\densf(\dd \coord_\dpwr)\:.
\end{equation}
\end{corollary}
\begin{IEEEproof}
We have
\begin{equation}
\expect_\ppp\left[\sum_{\someset\subseteq\ppp\atop|\someset|=\dpwr}\prod_{\pointi\in\someset}\h(\pointi,\ppp)\right]=\frac{1}{\dpwr!}\,\expect_\ppp\left[\usum_{\pointi\in\ppp^\dpwr}\prod_{\indi=1}^\dpwr\h(\pointi_\indi,\ppp)\right]\:,
\end{equation}
since the sum on the right hand side considers each product of the left hand side exactly $\dpwr!$ times due to the ordering of the elements and since the coordinates are distinct.
Applying Lemma~\ref{th:CMgen} yields the result.
\end{IEEEproof}

\subsection{Sum-Product Functionals on PPPs}
In this section we derive an expression for the expected value given in~\eqref{eq:goal}. The following lemma serves as preparation.
\begin{lemma}\label{lem2}\it
%\comment{Should we say that all elements in $\someset$ are different?}
Let $\someset\subseteq\re^\dimensions$ be a countable set and $\fcount\in\n$.
Let the vector of exponents $\fexp=(\fexp_1,\dots,\fexp_\fcount)\in\n^\fcount$ with $\|\fexp\|_1=\sum_{\indi=1}^\fcount \fexp_\indi$.
We assume $\|\fexp\|_1>0$.
Further, let $\fexpc(\indi)=\sum_{\indii=1}^\indi\fexp_\indii$ for all $\indi\in[\fcount]$ be the cumulative sum of the exponents $\fexp_\indi$; we set $\fexpc(0)=0$.
Let $\pointi=\big(\pointi^{(1)},\dots,\pointi^{(\fcount)}\big)\in\someset^{\|\fexp\|_1}$ with $\pointi^{(\indi)}=\big(\pointi_{\fexpc(\indi-1)+1},\dots,\pointi_{\fexpc(\indi)}\big)$. 
Finally, let $\f_\indi:\pppspace\rightarrow\re$ with $1\leq \indi\leq \fcount$ be non-negative functions.
Then we have
\begin{eqnarray}\label{lem2:main}
\eqleft{
\sum_{\pointi\in \someset^{\|\fexp\|_1}}\prod_{\indi=1}^\fcount\prod_{\indii=1}^{\fexp_\indi}\f_\indi\left(\pointi_{\indii}^{(\indi)}\right)
}\eqnln
\sum_{\indlen=1}^{\maxlens}
%\frac{1}{\indlen!}
\sum_{\mat\in\matset}
\magic_\mat\sum_{\somesetii\subseteq\someset\atop|\somesetii|=\indlen}\prod_{\indi=1}^\fcount\prod_{\indii=1}^\indlen\f_\indi^{\melem_{\indi\indii}}\big(\pointii_\indii\big)\:,
\end{eqnarray}
where $\matset\subseteq\n^{\fcount\times\indlen}$ is the class of all $\fcount\times\indlen$ matrices for which the columns $\|\melem_{\cdot\indii}\|_1>0$ for $\indii=[\indlen]$ and the rows $\|\melem_{\indi\cdot}\|_1=\fexp_\indi$ for all $\indi=[\fcount]$,
$\mat=(\melem_{\indi\indii})$ with $1\leq \indi\leq \fcount$ and $1\leq \indii\leq \indlen$, and $\somesetii=\{\pointii_1,\ldots,\pointii_\indlen\}$ without any specific ordering and the variable $\magic_\mat$ is defined as
\begin{equation}
\magic_\mat=\prod_{\indi=1}^\fcount\frac{\fexp_\indi!}{\prod_{\indii=1}^\indlen\melem_{\indi\indii}!}\:.
\end{equation}
%The symbol $\neq$ indicates that we sum only over $\pointi\in\someset^\indlen$ with distinct coordinates, i.e., $\pointi_\indiv\neq\pointi_\indv$ for all $0\leq\indiv<\indv\leq\indlen$.

%\comment{Should we use the notation $\fcum_\indi$ in the Lemma itself?}

%\comment{TODO: add a toy example for the lemma}

%\comment{Optimize by not dividing by $l!$ and ordering the distinct vector $u$ on the right hand side, maybe by the norm of points. I don't know how to do this!?}
\end{lemma}
%The left hand side product might contain equal values of $\f_\indii(\pointi_\indi)$. The purpose of this lemma is to avoid this by multiplying such elements only once and raising them to the corresponding powers~$\melem_{\indi\indii}$.
\begin{IEEEproof}
We prove the lemma by showing that the same products are summed on the left and on the right hand side. 
Note that all terms in each sum are non-negative and hence the sums on both sides are either absolutely convergent or diverge to infinity irrespective of the order of the summation; in both cases we can exchange the order of the summation.

We define the function $\isom(\pointi)=(\indlen,\somesetii,\mat)$, as follows:
$\indlen$ is the number of distinct coordinates of $\pointi$, i.e., $\indlen=|\{\pointi_\indi\,|\,1\leq \indi\leq \|\fexp\|_1\}|$. Clearly, $1\leq \indlen\leq \min(\|\fexp\|_1,|\someset|)$, covering the range from all $\pointi_\indi$ being the same to all $\pointi_\indi$ being different.
%The vector $\pointi$ consists of at most $|\someset|$ distinct elements.
The set $\somesetii$ is the set of all distinct elements in $\pointi$, i.e., $\somesetii=\{\pointi_\indi\,|\,1\leq \indi\leq \|\fexp\|_1\}$. Hence, $\somesetii\subseteq \someset$ with $|\somesetii|=\indlen$. We denote the elements of $\somesetii=\{\pointii_1,\dots,\pointii_\indlen\}$ without any specific ordering.
The matrix $\mat$ is of size $\fcount\times\indlen$ with entries $\melem_{\indi\indii}$, defined as follows: 
$\melem_{\indi\indii}$ is the number of coordinates in the vector $\pointi^{(\indi)}$ that are equal to $\pointii_\indii$, i.e., $\melem_{\indi\indii}=|\{\indiii\,|\,\pointi^{(\indi)}_\indiii=\pointii_\indii,1\leq\indiii\leq\fexp_\indi\}|$.
Note that these numbers must sum to $\sum_{\indii=1}^\indlen\melem_{\indi\indii}=\fexp_\indi$ for all $\indi\in[\fcount]$, and each element $\pointii_\indii$ must occur at least once in the product, i.e., $\sum_{\indi=1}^\fcount\melem_{\indi\indii}>0$ for all $\indii\in[\indlen]$.

The function $\isom(\pointi)$ is not injective, i.e., there can be different vectors $\pointi \neq \pointi'$ with $\isom(\pointi)=\isom(\pointi')$.
Let $(\indlen,\somesetii,\mat)$ be arbitrary, but fixed. In the following we calculate the size of the preimage $\isom^{-1}(\indlen,\somesetii,\mat)$. Toward this goal, let $\pointi\in\isom^{-1}(\indlen,\somesetii,\mat)$ be an arbitrary member of this preimage. Further, let us use the notation
\begin{equation}\label{eq:lem2:prod}
\fcum_i\big(\pointi^{(\indi)}\big)=\prod_{\indii=1}^{\fexp_\indi}\f_\indi\left(\pointi_\indii^{(\indi)}\right)\:.
\end{equation}
Observe that the product $\prod_{\indi=1}^\fcount\fcum_i\big(\pointi^{(\indi)}\big)$ is invariant to permutations inside the vectors $\pointi^{(\indi)}$. For each $\pointi^{(\indi)}$, the number of such permutations is $\fexp_\indi!$, but whether such permutations actually result in a different $\pointi^{(\indi)}$ (and thus $\pointi$) depends on how many distinct elements of $\someset$ appear in $\pointi^{(\indi)}$. This is determined by the $i$th row of $\mat$. Hence, the number of permutations resulting in a different $\pointi^{(\indi)}$ is
\begin{equation}
\cperm_\indi=\frac{\fexp_\indi!}{\prod_{\indii=1}^\indlen\melem_{\indi\indii}!}\:.
\end{equation}
The number of permutations of $\pointi$, which lead to the same products~\eqref{eq:lem2:prod} is hence $\magic_\mat=\prod_{\indi=1}^\fcount\cperm_\indi$.
%Since the order of the coordinates in $\pointii$ is based on their first appearence in $\pointi$, these permutations may lead to a permuted $\pointii$ and hence a different vector $(\indlen,\pointii,\mat)$. To account for this, we have to divide the number of permutations of $\pointi$ by the number of permutations of $\pointii$, which is given by $\indlen!$.
Hence, the preimage $\isom^{-1}(\indlen,\somesetii,\mat)$ of a given vector $(\indlen,\somesetii,\mat)$ contains $|\isom^{-1}(\indlen,\somesetii,\mat)|=\magic_\mat=\prod_{\indi=1}^\fcount\cperm_\indi$ elements.
Note that the union of all these preimages gives $\someset^{\|\fexp\|_1}$.

Each of the elements in a preimage leads to the same product on the left hand side of~\eqref{lem2:main}, i.e.,
\begin{equation}
\prod_{\indi=1}^\fcount\fcum_i\big(\pointi^{(\indi)}\big)=\prod_{\indi=1}^\fcount\prod_{\indii=1}^\indlen\f_\indi^{\melem_{\indi\indii}}\big(\pointii_\indii\big)\:.
\end{equation}
Hence, in the right hand side we sum over all combinations $(\indlen,\somesetii,\mat)$, for each combination multiplying one element of the corresponding preimage $\isom^{-1}(\indlen,\somesetii,\mat)$ by the size $\magic_\mat$ of this preimage.
%Finally, we have to divide the sum by $\indlen!$ to avoid taking into account the permutations of the vector $\pointii$ several times, since these permutations are already covered in $\magic_\mat$.
\end{IEEEproof}
%Now we show our main result. \comment{Replace text by text from Introduction}

Next, we state our main result on sum-product functionals on PPPs of the general form~\eqref{eq:goal}.
\begin{theorem}[Sum-product functionals for PPPs]\label{th6}\it
Let $\ppp$ be a PPP with a locally finite and diffuse intensity measure $\densf$. Also, let $\f_\indi,\g:\pppspace\times\rndvardom\rightarrow\re$ with $\rndvardom\subseteq\re$ and $1\leq\indi\leq\fcount$ be non-negative measurable functions with $\g(\coord)\leq 1$ for all $\coord\in\pppspace$ and $\fexp_\indi\in\n$ with $\|\fexp\|_1>0$. Furthermore, let $\rndvar_\cdot=(\rndvar_\pointi:\pointi\in\ppp)$ be a family of $\rndvardom$-valued i.i.d. random marks of the points in $\ppp$.
Then we have
\begin{eqnarray}\label{eq:th6}
\lefteqn{
\expect_{\ppp,\rndvar_\cdot}\left[\prod_{\indi=1}^\fcount\left(\sum_{\pointi\in\ppp}\f_\indi(\pointi,\rndvar_\pointi)\right)^{\fexp_\indi}\prod_{\pointii\in\ppp}\g(\pointii,\rndvar_\pointii)\right]}\\\nonumber
&=&
\exp\left(-\int_{\pppspace}\big(1-\expect_{\rndvar}\big[\g(\coord,\rndvar)\big]\big)\,\densf(\dd \coord)\right)
%\\\nonumber&&
\expect_{|\ppp|}\left[
\sum_{\indlen=1}^{\maxlen}
\sum_{\mat\in\matset}
\frac{\magic_\mat}{\indlen!}
\prod_{\indi=1}^\indlen \int_{\pppspace}\expect_\rndvar\Big[\g(\coord,\rndvar)\prod_{\indii=1}^\fcount\f_\indii^{\melem_{\indi\indii}}(\coord,\rndvar)\Big]\,\densf(\dd \coord)\right] \:,
\end{eqnarray}
where $\rndvar$ denotes a random variable with the same distribution as the i.i.d. random variables $\rndvar_\pointi$. Further, $\matset$ is the class of all $\fcount\times\indlen$ matrices with non-negative integer entries for which the columns $\|\melem_{\cdot\indii}\|_1>0$ for $\indii\in[\indlen]$ and the rows $\|\melem_{\indi\cdot}\|_1=\fexp_\indi$ for all $\indi\in[\fcount]$, and
$\magic_\mat=\prod_{\indiv=1}^\fcount\frac{\fexp_\indiv!}{\prod_{\indv=1}^\indlen\melem_{\indiv\indv}!}$.
Note that the expected value $\expect_{|\ppp|}$ can be omitted iff $\densf\big(\pppspace\big)=\infty$ a.s.
\end{theorem}
\begin{IEEEproof}
\begin{eqnarray}
\lefteqn{\expect_{\ppp,\rndvar_\cdot}\left[\prod_{\indi=1}^\fcount\left(\sum_{\pointi\in\ppp}\f_\indi(\pointi,\rndvar_\pointi)\right)^{\fexp_\indi}\prod_{\pointii\in\ppp}\g(\pointii,\rndvar_\pointii)\right]}\\\nonumber
&\stackrel{(a)}{=}&
\expect_{\ppp,\rndvar_\cdot}\left[\sum_{\pointi\in\ppp^{\|\fexp\|_1}}\prod_{\indi=1}^\fcount\prod_{\indii=1}^{\fexp_\indi}\f_\indi\left(\pointi_{\indii}^{(\indi)},\rndvar_{\pointi_{\indii}^{(\indi)}}\right)\prod_{\pointii\in\ppp}\g(\pointii,\rndvar_\pointii)\right]\\\nonumber
&\stackrel{(b)}{=}&
\expect_{\ppp,\rndvar_\cdot}\left[\sum_{\indlen=1}^{\maxlen}
%\frac{1}{\indlen!}
\sum_{\mat\in\matset}
\magic_\mat\sum_{\someset\subseteq\ppp\atop|\someset|=\indlen}\prod_{\indi=1}^\fcount\prod_{\indii=1}^\indlen\f_\indi^{\melem_{\indi\indii}}\big(\pointi_\indii,\rndvar_{\pointi_\indii}\big)\prod_{\pointii\in\ppp}\g\big(\pointii,\rndvar_\pointii\big)\right]\\\nonumber
&\stackrel{(c)}{=}&
\expect_\ppppc\left[
\sum_{\indlen=1}^{\maxlenk}
%\frac{1}{\indlen!}
\sum_{\mat\in\matset}
\magic_\mat\expect_{\ppp,\rndvar_\cdot}\left[\sum_{\someset\subseteq\ppp\atop|\someset|=\indlen}\prod_{\indi=1}^\fcount\prod_{\indii=1}^\indlen\f_\indi^{\melem_{\indi\indii}}\big(\pointi_\indii,\rndvar_{\pointi_\indii}\big)\prod_{\pointii\in\ppp}\g\big(\pointii,\rndvar_\pointii\big)\,\,\bigg|\,\,|\ppp|=\ppppc\right]\right]\\\nonumber
&\stackrel{(d)}{=}&
\expect_\ppppc\left[
\sum_{\indlen=1}^{\maxlenk}
%\frac{1}{\indlen!}
\sum_{\mat\in\matset}
\magic_\mat\expect_{\ppp,\rndvar_\cdot}\left[\sum_{\someset\subseteq\ppp\atop|\someset|=\indlen}\prod_{\indii=1}^\indlen \g\big(\pointi_\indii,\rndvar_{\pointi_\indii}\big)\prod_{\indi=1}^\fcount\f_\indi^{\melem_{\indi\indii}}\big(\pointi_\indii,\rndvar_{\pointi_\indii}\big)\prod_{\pointii\in\ppp\backslash\{\pointi_\indiii\}_{\indiii=1}^\indlen}\g(\pointii,\rndvar_\pointii)\,\,\bigg|\,\,|\ppp|=\ppppc\right]\right]\\\nonumber
&\stackrel{(e)}{=}&
\expect_\ppppc\left[
\sum_{\indlen=1}^{\maxlenk}
%\frac{1}{\indlen!}
\sum_{\mat\in\matset}
\magic_\mat\expect_{\ppp}\left[\sum_{\someset\subseteq\ppp\atop|\someset|=\indlen}\prod_{\indii=1}^\indlen \expect_{\rndvar_\cdot}\left[\g\big(\pointi_\indii,\rndvar_{\pointi_\indii}\big)\prod_{\indi=1}^\fcount\f_\indi^{\melem_{\indi\indii}}\big(\pointi_\indii,\rndvar_{\pointi_\indii}\big)\right]\prod_{\pointii\in\ppp\backslash\{\pointi_\indiii\}_{\indiii=1}^\indlen}\expect_{\rndvar_\cdot}\left[\g(\pointii,\rndvar_\pointii)\right]\,\,\bigg|\,\,|\ppp|=\ppppc\right]\right]\\\nonumber
&\stackrel{(f)}{=}&
\expect_\ppppc\left[
\sum_{\indlen=1}^{\maxlenk}\frac{1}{\indlen!}
\sum_{\mat\in\matset}
\magic_\mat
\int_\pppspace\dots\int_\pppspace\int_\pppnk
\prod_{\indii=1}^\indlen \expect_{\rndvar_\cdot}\left[\g\big(\coord_\indii,\rndvar_{\indii}\big)\prod_{\indi=1}^\fcount\f_\indi^{\melem_{\indi\indii}}\big(\coord_\indii,\rndvar_{\indii}\big)\right]
\right.\\\nonumber&&\left.\hspace{1cm}\cdot\,
\prod_{\pointii\in\pppre\backslash\{\coord_\indiii\}_{\indiii=1}^\indlen}\expect_{\rndvar_\cdot}\left[\g(\pointii,\rndvar_\pointii)\right]
\pppp_{\{\coord_\indiii\}_{\indiii=1}^\indlen}(\dd \pppre)
\densf(\dd \coord_1)\cdots\densf(\dd \coord_\indlen)\right]\\\nonumber
&\stackrel{(g)}{=}&
\expect_\ppppc\left[
\sum_{\indlen=1}^{\maxlenk}\frac{1}{\indlen!}
\sum_{\mat\in\matset}
\magic_\mat
\int_\pppspace\dots\int_\pppspace
\prod_{\indii=1}^\indlen \expect_{\rndvar_\cdot}\left[\g\big(\coord_\indii,\rndvar_{\indii}\big)\prod_{\indi=1}^\fcount\f_\indi^{\melem_{\indi\indii}}\big(\indii,\rndvar_{\coord_\indii}\big)\right]
\right.\\\nonumber&&\left.\hspace{1cm}\cdot\,
\int_\pppnk \prod_{\pointii\in\pppre}\expect_{\rndvar_\cdot}\left[\g(\pointii,\rndvar_\pointii)\right]
\pppp(\dd \pppre)
\densf(\dd \coord_1)\cdots\densf(\dd \coord_\indlen)\right]\\\nonumber
&=&
\expect_\ppppc\left[
\sum_{\indlen=1}^{\maxlenk}\frac{1}{\indlen!}
\sum_{\mat\in\matset}
\magic_\mat
\prod_{\indii=1}^\indlen
\int_\pppspace
\expect_\rndvar\left[\g\big(\coord,\rndvar\big)\prod_{\indi=1}^\fcount\f_\indi^{\melem_{\indi\indii}}\big(\coord,\rndvar\big)\right]\expect_\ppp\left[\prod_{\pointii\in\ppp}\expect_\rndvar\left[\g(\pointii,\rndvar_\pointii)\right]
\,\,\bigg|\,\,|\ppp|=\ppppc\right]
\densf(\dd \coord)\right]\:.
\end{eqnarray}
In $(a)$ we apply Lemma~\ref{lem:exchange} presented in Appendix~\ref{sec:app:lemmata}, where $\someset={\indii}^{(\indi)}$ is defined as in Lemma~\ref{lem2}. In $(b)$ we have $\pointi=\{\pointi_1,\dots,\pointi_{\|\fexp\|_1}\}$ and Lemma~\ref{lem2} is applied; in $(c)$ we condition on the number of points $|\ppp|$; in $(d)$ we factor out all $\g(\pointii)$ from the rightmost product for which $\pointii=\pointi_\indii$ for some $\indii$. This is possible since all $\pointi_\indii$ are distinct (This property made necessary the use of Lemma 2). In $(e)$ we move the expected value of the random variables $\rndvar$ inside the sum/product since the $\rndvar$ are i.i.d. In $(f)$, $\rndvar$ and $\rndvar_\indii$ denote i.i.d. random variables with the same distribution as the marks $\rndvar_\pointi$ of $\ppp$. Further, we apply Corollary~\ref{cor:CMgen}, and $\pppp_{\{\coord_\indiii\}_{\indiii=1}^\indlen}$ denotes the Palm distribution of $\ppp$ conditioning on the points $\coord_\indiii$, $\indiii=1,\dots,\indlen$. The symbol $\pppnk$ denotes the space of counting measures with at most $\ppppc$ points in all Borel sets $B$.
Note that $\pppnk\subseteq\pppn$. $(g)$ holds due to Slivnyak's theorem.

Calculating the probability generating functional according to \eqref{eq:pgfl} yields the result.

%\comment{OPEN ISSUE: The expression $\min(\|p\|_1,|\ppp|)$ is random!}
%\comment{Step (e): Is there a theorem to cite for $\int_\pppn\prod_{u\in\phi\backslash\{x\}} P_x(\dd \phi)=\int_\pppn\prod_{u\in\phi} P(\dd \phi)$?}
\end{IEEEproof}
Please note that in Theorem~\ref{th6} there is no requirement that the $f_i$ are different. Hence, without loss of generality we could set all $p_i=1$ and $q=\|p\|_1$ instead.
We decided, however, to present the more general case, since it reduces the number of terms that have to be evaluated on the right hand side due to the powers $\melem_{\indi\indii}$.

\subsection{Some Special Cases}
We now highlight some special cases of Theorem~\ref{th6}, which are of interest for the following sections.
\begin{corollary}[Stationary PPPs]\label{cor:stationary}\it
%Let $\ppp$ be a stationary Poisson point process with intensity $\dens$. Let further $\f_\indi,\g:\pppspace\times\rndvardom\rightarrow\re$  be non-negative measurable functions with $\g(\coord)\leq 1$ for all $\coord\in\pppspace$ and $\fexp_\indi\in\n$ (we assume $0\in\n$) with $\fexp_\indi>0$ for $1\leq\indi\leq\fcount$. Furthermore, let $\rndvar=(\rndvar_\coord)$ for all $\coord\in\pppspace$ be a family of $\rndvardom$-valued i.i.d. random variables.
When the PPP is stationary with intensity $\dens$, under the same assumptions as in Theorem~\ref{th6}, we have
\begin{eqnarray}
\lefteqn{
\expect_{\ppp,\rndvar}\left[\prod_{\indi=1}^\fcount\left(\sum_{\pointi\in\ppp}\f_\indi(\pointi,\rndvar_\pointi)\right)^{\fexp_\indi}\prod_{\pointi\in\ppp}\g(\pointi,\rndvar_\pointi)\right]}\\\nonumber
&=&
\exp\left(-\dens\int_{\pppspace}\big(1-\expect_{\rndvar}\big[\g(\coord,\rndvar_\coord)\big]\big)\,\dd \coord\right)
%\\\nonumber&&
\sum_{\indlen=1}^{\maxlenhomo}
\sum_{\mat\in\matset}
\frac{\magic_\mat}{\indlen!}
\prod_{\indi=1}^\indlen \dens\int_{\pppspace}\expect_\rndvar\Big[\g(\coord,\rndvar_\coord)\prod_{\indii=1}^\fcount\f_\indii^{\melem_{\indi\indii}}(\coord,\rndvar_\coord)\Big]\,\dd \coord\:,
\end{eqnarray}
where $\matset$ is the class of all $\fcount\times\indlen$ matrices with non-negative integer entries for which the columns $\|\melem_{\cdot\indii}\|_1>0$ for $\indii\in[\indlen]$ and the rows $\|\melem_{\indi\cdot}\|_1=\fexp_\indi$ for $\indi\in[\fcount]$, and
$\magic_\mat=\prod_{\indiv=1}^\fcount\frac{\fexp_\indiv!}{\prod_{\indv=1}^\indlen\melem_{\indv\indiv}!}$.
\end{corollary}
\begin{IEEEproof}
Substituting $\densf(\someset)=\dens\meas(\someset)$ for all Borel sets $\someset\subseteq\pppspace$ in Theorem~\ref{th6} yields the result.
\end{IEEEproof}

\begin{corollary}[Stationary PPPs with $\fcount=1$]\label{cor:singlef}\it
%Let $\ppp$ be a stationary Poisson point process with intensity $\dens$. Let further $\f,\g:\pppspace\times\rndvardom\rightarrow\re$  be non-negative measurable functions with $\g(\coord)\leq 1$ for all $\coord\in\pppspace$ and $\fexp\in\n$ (we assume $0\in\n$) with $\fexp>0$. Furthermore, let $\rndvar=(\rndvar_\coord)$ for all $\coord\in\pppspace$ be a family of $\rndvardom$-valued i.i.d. random variables.
When the PPP is stationary with $\fcount=1$, under the same assumptions as in Theorem~\ref{th6}, we have
\begin{eqnarray}
\lefteqn{
\expect_{\ppp,\rndvar}\left[\left(\sum_{\pointi\in\ppp}\f(\pointi,\rndvar_\pointi)\right)^{\fexp}\prod_{\pointi\in\ppp}\g(\pointi,\rndvar_\pointi)\right]}\\\nonumber
&=&
\exp\left(-\dens\int_{\pppspace}\big(1-\expect_{\rndvar}\big[\g(\coord,\rndvar_\coord)\big]\big)\,\dd \coord\right)
%\\\nonumber&&
\sum_{\indlen=1}^{\fexp}
\sum_{\mat\in\matset}
\frac{\magic_\mat}{\indlen!}
\prod_{\indi=1}^\indlen \dens\int_{\pppspace}\expect_\rndvar\Big[\g(\coord,\rndvar_\coord)\f^{\melem_{\indi}}(\coord,\rndvar_\coord)\Big]\,\dd \coord\:,
\end{eqnarray}
where $\matset$ is the class of all vectors of length $\indlen$ having strictly positive integer coordinates which sum up to $\fexp$, and
$\magic_\mat=\frac{\fexp!}{\prod_{\indv=1}^\indlen\melem_{\indv}!}$.
\end{corollary}
\begin{IEEEproof}
Substituting $\fcount=1$ in Corollary~\ref{cor:stationary} yields the result.
\end{IEEEproof}

\begin{corollary}[Stationary PPPs with $\fcount=1$ and $\fexp=1$]\label{cor:nopower}\it
%Let $\ppp$ be a stationary Poisson point process with intensity $\dens$. Let further $\f,\g:\pppspace\times\rndvardom\rightarrow\re$  be non-negative measurable functions with $\g(\coord)\leq 1$ for all $\coord\in\pppspace$. Furthermore, let $\rndvar=(\rndvar_\coord)$ for all $\coord\in\pppspace$ be a family of $\rndvardom$-valued i.i.d. random variables.
When the PPP is stationary and $\fcount=1$, $\fexp=1$, under the same assumptions as in Theorem~\ref{th6}, we have
\begin{equation}
\expect_{\ppp,\rndvar}\left[\sum_{\pointi\in\ppp}\f(\pointi,\rndvar_\pointi)\prod_{\pointi\in\ppp}\g(\pointi,\rndvar_\pointi)\right]=
\exp\left(-\dens\int_{\pppspace}\big(1-\expect_{\rndvar}\big[\g(\coord,\rndvar_\coord)\big]\big)\,\dd \coord\right)
\dens\int_{\pppspace}\expect_\rndvar\Big[\f(\coord,\rndvar_\coord)\g(\coord,\rndvar_\coord)\Big]\,\dd \coord\:.
\end{equation}
\end{corollary}
\begin{IEEEproof}
Substituting $\fexp=1$ in Corollary~\ref{cor:singlef} implies that $\matset=\{(1)\}$ and $\magic_\mat=1$.
\end{IEEEproof}
Note that Corollary~\ref{cor:nopower} also follows from the Campbell-Mecke theorem~\cite{haenggi13:book} as given in~\eqref{eq:mecke}.

\section{Interference Functionals}\label{sec:expect}
%In this section we want to calculate expected values of the form $\expect\big[\interf_\indi^{\fexp_\indi}\interf_\indii^{\fexp\indii}\,\exp\big(-\const (\interf_\indi+\interf_\indii)\big)\big]$ for general path-loss and fading models and then show some special cases for commonly used models.

\subsection{Modeling Assumptions}
We consider a wireless network with interferers distributed according to a PPP $\ppp$ with a locally finite and diffuse intensity measure $\densf$.
%The symbol $\pointi\in\ppp$ denotes both the node and its location for simplicity of notation, by. 
All interferers transmit with the same transmission power, which we set to one. Time is slotted, and slotted ALOHA is employed for medium access control, i.e., each node in $\ppp$ accesses the channel in each time slot independently with a certain probability $\txp$. Let $\ppp_\indi\subseteq\ppp$ denote the set of interferers that are active in slot $\indi$. The interference power received at the origin $\origin$ in slot $\indi$ is modeled by
\begin{equation}\label{eq:interf}
\interf_\indi=\sum_{\pointi\in\ppp}\fgain_\indi(\pointi)\gain(\pointi)\indici\:.
\end{equation}
Here, $\indicator$ denotes the indicator function, i.e., 
\begin{equation}
\indici=
\begin{cases}
1,&\pointi\in\ppp_\indi,\\
0,&\mbox{else.}
\end{cases}
\end{equation}
The term $\gain(\pointi)$ denotes the path gain from node $\pointi$ to $\origin$, which is assumed to be a non-negative function that decreases monotonically with $\|\pointi\|_2$ with $\lim_{\|\pointi\|_2\rightarrow\infty}\gain(\pointi)=0$. Table~\ref{tab:pathloss} summarizes some commonly used models for the path gain, where $\pointi\in\ppp$ is an arbitrary point, $\plexp>2$ is the path loss exponent, and $\pleps>0$. The fading coefficient $\fgain_\indi(\pointi)$ denotes the channel fading state, i.e., $\fgain_\indi(\pointi)$ is a random variable that follows some distribution that depends on the fading model. Note that $\fgain_\indi(\pointi)$ are i.i.d.~for different points $\pointi$ or different time slots $\indi$. We assume that the fading coefficients have an expected value $\expect[\fgain_\indi(\pointi)]=1$. Table~\ref{tab:fading} shows various well-known fading models. The symbol $\BesselI_k(x)$ denotes the second modified Bessel function.

\begin{table}[!h]
\caption{Path gain models}\label{tab:pathloss}
\begin{center}
\renewcommand{\arraystretch}{1.5}
\begin{tabular}{l|l}
Model & Expression\\\hline
Singular model & $\gains=\|\pointi\|_2^{-\plexp}$\\
Minimum model & $\gainm=\min(1,\|\pointi\|_2^{-\plexp})$\\
$\pleps$ model & $\gaine=\frac{1}{\pleps+\|\pointi\|_2^{\plexp}}$\\
Distance+1 model & $\gaind=\frac{1}{(1+\|\pointi\|_2)^{\plexp}}$
\end{tabular}
\renewcommand{\arraystretch}{1}
\end{center}
\end{table}

\begin{table}[!h]
\caption{Fading models}\label{tab:fading}
\begin{center}
\renewcommand{\arraystretch}{1.5}
\begin{tabular}{l|l}
Fading model & Probability density function of the power, $x\geq 0$\\\hline
Rayleigh & $\pdf_\fgain(\coord)=\exp(-\coord)$\\
Erlang & $\pdf_\fgain(\coord)=\coord^{\fadk-1}\frac{\exp(-\coord)}{(\fadk-1)!}$ for $\fadk\in\n\backslash\{0\}$\\
Rice & $\pdf_\fgain(\coord)=\exp(-(\coord+\rilam)/2)\left(\frac{\coord}{\rilam}\right)^{\frac{\fadk}{4}-\frac{1}{2}}\frac{\BesselI_{\frac{\fadk}{2}-1}(\sqrt{\rilam \coord})}{2}$\\
& for $\fadk\in\n$ and $\rilam\in\re^+$\\
Nakagami & $\pdf_\fgain(\coord)=\coord^{\nakagamim-1}\frac{\exp(-\coord \nakagamim)}{\Gamma(\nakagamim)}\nakagamim^\nakagamim$ for $\nakagamim\in\re^+$\\
\end{tabular}
\renewcommand{\arraystretch}{1}
\end{center}
\end{table}

\subsection{The General Case}
In the following we derive an expression for the expected value $\expect\left[\prod_{\indi=1}^\fcount\interf_\indi^{\fexp_\indi}\,\exp\big(-\const \interf_\indi\big)\right]$ with general models for fading and path loss.
This is an important result for analyzing interference in wireless networks, as it occurs in the derivation of transmission success probabilities in many scenarios. An example is the success probability in Poisson networks with Nakagami fading, as is presented in Section~\ref{sec:nakagami}.
\begin{theorem}[Interference functionals]\label{th:expectmain}\it
Let $c\in\re$ be a constant, $\fexp=(\fexp_1,\dots,\fexp_\fcount)\in\n^\fcount$ with $\|\fexp\|_1>0$, and let $\interf_\indi$ denote the interference at the origin $\origin$ at time slot $\indi$ as defined in~\eqref{eq:interf}. Then we have
\begin{eqnarray}\label{eq:expectmain}
\lefteqn{\expect_{\ppp,\fgain,\indicator}\left[\prod_{\indi=1}^\fcount\interf_\indi^{\fexp_\indi}\,\exp\big(-\const \interf_\indi\big)\right]}\\\nonumber
&=&\exp\left(-\int_{\pppspace}\bigg(1-\prod_{\indii=1}^\fcount\Big(\txp\,\expect_{\fgain_\indii(\coord)}\left[\exp\big(-\const\fgain_\indii(\coord)\gain(\coord)\big)\right]+1-\txp\Big)\bigg)\,\densf(\dd \coord)\right)
%\\\nonumber&&
\expect_{|\ppp|}\Bigg[
\sum_{\indlen=1}^{\maxlen}
\sum_{\mat\in\matset}
\frac{\magic_\mat}{\indlen!}
\\\nonumber&&
\prod_{\indi=1}^\indlen
\int_{\pppspace}
\prod_{\indii=1}^\fcount\left(\txp\,\expect_{\fgain_\indii(\coord)}\left[\exp\big(-\const\fgain_\indii(\coord)\gain(\coord)\big)\big(\fgain_\indii(\coord)\gain(\coord)\big)^{\melem_{\indi\indii}}\right]+(1-\txp)\,\indim\right)\,\densf(\dd \coord)\Bigg]\:,
\end{eqnarray}
where $\matset$ is the class of all $\fcount\times\indlen$ matrices with non-negative integer entries for which the columns $\|\col_\indii\|_1>0$ for $\indii=1,\dots,\indlen$ and the rows $\|\row_\indi\|_1=\fexp_\indi$ for all $\indi=1,\dots,\fcount$.
\end{theorem}
Note that the term $(1-\txp)\,\indim$ can be omitted if $\fcount=1$, since all exponents $\melem_{\indi\indii}>0$.
\begin{IEEEproof}
The proof is based on Theorem~\ref{th6}. We have
\begin{eqnarray}
\lefteqn{\expect_{\ppp,\fgain,\indicator}\left[\prod_{\indi=1}^\fcount\interf_\indi^{\fexp_\indi}\,\exp\big(-\const \interf_\indi\big)\right]
}\\\nonumber
&=&\expect_{\ppp,\fgain,\indicator}\left[\prod_{\indi=1}^\fcount\interf_\indi^{\fexp_\indi}\prod_{\indii=1}^\fcount\exp\big(-\const \interf_\indii\big)\right]\\\nonumber
&=&\expect_{\ppp,\fgain,\indicator}\left[\prod_{\indi=1}^\fcount\interf_\indi^{\fexp_\indi}\,\exp\Bigg(-\const \sum_{\indii=1}^\fcount\interf_\indii\Bigg)\right]\\\nonumber
&=&\expect_{\ppp,\fgain,\indicator}\left[\prod_{\indi=1}^\fcount\Bigg(\sum_{\pointi\in\ppp}\fgain_\indi(\pointi)\gain_\indi(\pointi)\indici\Bigg)^{\fexp_\indi}\,\exp\Bigg(-\const \sum_{\indii=1}^\fcount\sum_{\pointii\in\ppp}\fgain_\indii(\pointii)\gain_\indii(\pointii)\indicii\Bigg)\right]\\\nonumber
&=&\expect_{\ppp,\fgain,\indicator}\left[\prod_{\indi=1}^\fcount\Bigg(\sum_{\pointi\in\ppp}\fgain_\indi(\pointi)\gain_\indi(\pointi)\indici\Bigg)^{\fexp_\indi}\,\prod_{\pointii\in\ppp}\exp\Bigg(-\const \sum_{\indii=1}^\fcount\fgain_\indii(\pointii)\gain_\indii(\pointii)\indicii\Bigg)\right]\\\nonumber
&\stackrel{(a)}{=}&\exp\left(-\int_{\pppspace}\left(1-\expect_{\fgain(\coord),\indicator}\left[\exp\left(-\const\sum_{\indii=1}^\fcount\fgain_\indii(\coord)\gain_\indii(\coord)\indicxii\right)\right]\right)\,\densf(\dd \coord)\right)
%\\\nonumber&&
\expect_{|\ppp|}\Bigg[
\sum_{\indlen=1}^{\maxlen}
\sum_{\mat\in\matset}
\frac{\magic_\mat}{\indlen!}
\\\nonumber&&\hspace{5mm}
\prod_{\indi=1}^\indlen \int_{\pppspace}\expect_{\fgain(\coord),\indicator}\left[\exp\left(-\const\sum_{\indii=1}^\fcount\fgain_\indii(\coord)\gain_\indii(\coord)\indicxii\right)\prod_{\indiii=1}^\fcount\big(\fgain_\indiii(\coord)\gain_\indiii(\coord)\indicxiii\big)^{\melem_{\indi\indiii}}\right]\,\densf(\dd \coord)\Bigg]\\\nonumber
&\stackrel{(b)}{=}&\exp\left(-\int_{\pppspace}\bigg(1-\prod_{\indii=1}^\fcount\Big(\txp\,\expect_{\fgain_\indii(\coord)}\left[\exp\big(-\const\fgain_\indii(\coord)\gain(\coord)\big)\right]+1-\txp\Big)\bigg)\,\densf(\dd \coord)\right)
%\\\nonumber&&
\expect_{|\ppp|}\Bigg[
\sum_{\indlen=1}^{\maxlen}
\sum_{\mat\in\matset}
\frac{\magic_\mat}{\indlen!}
\\\nonumber&&\hspace{5mm}
\prod_{\indi=1}^\indlen
\int_{\pppspace}
\prod_{\indii=1}^\fcount\left(\txp\,\expect_{\fgain_\indii(\coord)}\left[\exp\big(-\const\fgain_\indii(\coord)\gain(\coord)\big)\big(\fgain_\indii(\coord)\gain(\coord)\big)^{\melem_{\indi\indii}}\right]+(1-\txp)\,\indim\right)\,\densf(\dd \coord)\Bigg]\:,
\end{eqnarray}
where $(a)$ holds due to Theorem~\ref{th6} with substituting $\f_\indi\big(\pointi,\fgain_\indi(\pointi)\big)=\fgain_\indi(\pointi)\gain_\indi(\pointi)$, $\g\big(\pointi,\fgain(\pointii)\big)=\exp\left(-\const\sum_{\indii=1}^\fcount\fgain_\indii(\pointii)\gain_\indii(\pointii)\right)$, and $\fgain(\pointii)=\big(\fgain_1(\pointii),\dots,\fgain_\fcount(\pointii)\big)$. The terms $\indicxii$ denote Bernoulli random variables with $\expect\big[\indicxii\big]=\txp$.
$(b)$ holds due to the independence of the fading coefficients $\fgain_\indii(\coord)$ and $\indicxii$.
\end{IEEEproof}
Note that, similar to Theorem~\ref{th6}, we can omit $\expect_{|\ppp|}$ iff $\densf\big(\pppspace\big)=\infty$ a.s.

\subsection{Case Studies: Results for Specific Fading and Path Loss Models}

%\subsection{Calculations}
In the following we calculate the expected value $\expect_{\ppp,\fgain,\indicator}[\interf\exp(-\interf)]$ for a stationary PPP $\ppp$ with intensity $\dens$.
This expression has to be evaluated when analyzing the outage probability in certain scenarios, e.g., in the case of Nakagami fading with $\nakagamim=2$.
We have
\begin{eqnarray}\label{eq:i_expi_gen}
\expect_{\ppp,\fgain,\indicator}[\interf\exp(-\interf)]
=\exp\left(-\txp\dens\int_{\pppspace}\Big(1-\expect_{\fgain(\coord)}\big[\exp\big(-\fgain(\coord)\gain(\coord)\big)\big]\Big)\,\dd \coord\right)\,\txp\dens\int_{\pppspace}\expect_{\fgain(\coord)}[\fgain(\coord)\gain(\coord)\exp\big(-\fgain(\coord)\gain(\coord)\big)]\,\dd \coord
\end{eqnarray}
as a special case of Theorem~\ref{th:expectmain} with $\fcount=1$, $\fexp_1=1$, and $\const=1$.

\begin{table}[tb]
\begin{center}
\renewcommand{\arraystretch}{1.5}
\caption{Expected values for different fading models. Values can be substituted into~\eqref{eq:i_expi_gen}.}\label{tab:fadingexpect}
\begin{tabular}{l|cc}
Fading model & $\expect_{\fgain(\coord)}[\exp\big(-\!\fgain(\coord)\gain(\coord)\big)]$ & $\expect_{\fgain(\coord)}[\fgain(\coord)\gain(\coord)\exp\big(-\!\fgain(\coord)\gain(\coord)\big)]$\\\hline
Rayleigh & $\frac{1}{1+\gain(\coord)}$ & $\frac{\gain(\coord)}{(1+\gain(\coord))^2}$\\
Erlang with $k=2$ & $\frac{1}{(1+\gain(\coord))^2}$ & $\frac{2\gain(\coord)}{(1+\gain(\coord))^3}$\\
Erlang & $\frac{1}{(1+\gain(\coord))^k}$ & $\frac{k\gain(\coord)}{(1+\gain(\coord))^{k+1}}$\\
Rice & $\exp\left(-\frac{\psi\gain(\coord)}{1+2\gain(\coord)}\right)\big(1+2\gain(\coord)\big)^{-\frac{k}{2}}$ & $\exp\left(-\frac{\psi\gain(\coord)}{1+2\gain(\coord)}\right)\gain(\coord)\frac{k+\psi+2k\gain(\coord)}{\big(1+2\gain(\coord)\big)^{2+\frac{k}{2}}}$\\
Nakagami & $\left(\frac{\nakagamim}{\nakagamim+\gain(\coord)}\right)^\nakagamim$ & $\gain(\coord)\left(\frac{\nakagamim}{\nakagamim+\gain(\coord)}\right)^{\nakagamim+1}$
\end{tabular}
\textbf{\renewcommand{\arraystretch}{1}}
\end{center}
\end{table}

The expected values within the integrals depend on the specific fading model. The results for the fading models in Table~\ref{tab:fading} and the singular path gain are presented in Table~\ref{tab:fadingexpect}.
Substituting these expressions yields the final results.

As an example, for Rayleigh fading, the singular path gain, and a two-dimensional stationary PPP $\ppp$ with intensity $\dens$ we have
\begin{eqnarray}\label{eq:i_expi}
\expect_{\ppp,\fgain,\indicator}[\interf\exp(-\interf)]
%\\\nonumber
&=& \txp\dens\int_{\re^2}\frac{\gain(\coord)}{(1+\gain(\coord))^2}\,\dd \coord \exp\left(-\txp\dens\int_{\re^2}\left(1-\frac{1}{1+\gain(\coord)}\right)\,\dd \coord\right)\\\nonumber
&=& \falpha\scut\exp\left(-\scut\right)\:,
\end{eqnarray}
where $\falpha=\frac{2}{\plexp}$, $\scut=\frac{\txp\dens\pi}{\sinc(\falpha)}$, and $\sinc(x)=\frac{\sin(\pi x)}{\pi x}$.
Fig.~\ref{fig:E:IexpI} shows a plot of~\eqref{eq:i_expi} over the intensity $\dens$ for different path loss exponents $\plexp$ with $\txp=1$. As can be seen, the expected value possesses a peak at a certain density of interferers, which depends on $\plexp$. For increasing densities the expected value approaches zero, i.e., $\lim_{\dens\to\infty}\expect[\interf\exp(-\interf)]=0$ for all $\plexp>2$.

\begin{figure}[tbh]
\centering
\begin{tikzpicture}[scale=1]
\begin{axis}[xlabel={Intensity $\dens$},
	ylabel={$\expect[\interf\exp(-\interf)]$},ymin=0,ymax=0.3,xmin=0,xmax=0.8,grid=both,xtick={0,0.2,0.4,0.6,0.8,1},
            ytick={0,0.1,0.2,0.3},
xticklabels={0,0.2,0.4,0.6,0.8,1},
yticklabels={0,0.1,0.2,0.3},
legend style={at={(0.98,0.98)}, anchor=north east, font=\footnotesize},
legend cell align=left
]

\addplot plot[color=black,solid,no marks,style=thick]  table[x index=0,y index=1]  {E_I_expI.txt};\addlegendentry{ ~$\alpha=2.5$ }
\addplot plot[color=black,dashed,no marks,style=thick]  table[x index=0,y index=2]  {E_I_expI.txt};\addlegendentry{ ~$\alpha=3$ }
%\addplot plot[color=black,dotted,no marks,style=thick]  table[x index=0,y index=3]  {E_I_expI.txt};\addlegendentry{ ~$\alpha=3.5$ }
\addplot plot[color=black,densely dotted,no marks,style=thick]  table[x index=0,y index=4]  {E_I_expI.txt};\addlegendentry{ ~$\alpha=4$ }
%\addplot plot[color=black,densely dashed,no marks,style=thick]  table[x index=0,y index=5]  {E_I_expI.txt};\addlegendentry{ ~$\alpha=4.5$ }
\addplot plot[color=black,dashdotted,no marks,style=thick]  table[x index=0,y index=6]  {E_I_expI.txt};\addlegendentry{ ~$\alpha=5$ }

\end{axis}
\end{tikzpicture}
\caption{Expected value $\expect[\interf\exp(-\interf)]$ given in~\eqref{eq:i_expi} over the intensity of interferers for different path loss exponents $\plexp$ with $\txp=1$.}\label{fig:E:IexpI}
\end{figure}

%\comment{Add a plot: Three curves over $\lambda$ with $\alpha=2,3,4$ as parameter (CB).}

For the special case of $\alpha=4$ (i.e., $\falpha=\frac{1}{2}$) we get
\begin{equation}\label{eq:i_expi_al_iv}
\expect_{\ppp,\fgain,\indicator}[\interf\exp(-\interf)]=\frac{1}{4}\txp\dens\pi^2\exp\left(-\frac{\txp\dens\pi^2}{2}\right)\:,
\end{equation}
which coincides with the result calculated using the pdf of the interference (Equation~(3.22) in~\cite{haenggi09:interference}).

Next, we generalize this result by computing the expected value $\expect[I^k\exp(-I)]$ for $\indiii\in\n$.
% \comment{To Stavros: this result holds for all $k\geq 0$ and can be easily generalized to $\expect[\interf^\indiii\exp(-\dens \interf)]$.} 
As an intermediate result, for $\indi\in\n$, the expected value $\expect_{\fgain(\pointi)}[(\fgain(\pointi)\gain(\pointi))^\indi\exp\big(-\fgain(\pointi)\gain(\pointi)\big)]$ is given by
\begin{equation}\label{eq:ehikexp}
\expect_{\fgain(\pointi)}\big[\big(\fgain(\pointi)\gain(\pointi)\big)^\indi\exp\big(-\fgain(\pointi)\gain(\pointi)\big)\big]=\frac{\indi!\, \gain^\indi(\pointi)}{\big(1+\gain(\pointi)\big)^{\indi+1}}\:.
\end{equation}
Substituting this expression into~\eqref{eq:expectmain} allows the calculation of $\expect_{\ppp,\fgain(\pointi)}[\interf^\indiii\exp(-\interf)]$.
Some example expressions for $\indiii=2,3,4$ are presented in \eqref{eq:i2exp}, \eqref{eq:i3exp}, and~\eqref{eq:i4exp}, respectively. Here, again $\falpha=\frac{2}{\plexp}$, $\scut=\frac{\txp\dens\pi}{\sinc(\falpha)}$, and $\sinc(x)=\frac{\sin(\pi x)}{\pi x}$.
\begin{figure*}[!t]
\begin{align}\label{eq:i2exp}
\expect[I^2\exp(-I)]&=\exp\left(-\scut\right)\scut\big(\falpha(1-\falpha)+2\scut\falpha\big)\:.\\
\expect[I^3\exp(-I)]&=\exp\left(-\scut\right)\scut\,\frac{\falpha^2(\alpha-1)(\alpha-2)+2\scut\left(6(\alpha-2)+4\scut\right)}{\alpha}\:.\label{eq:i3exp}\\\label{eq:i4exp}
\expect[I^4\exp(-I)]&=\exp\left(-\scut\right)\scut\,\frac{4\alpha(\alpha-1)(\alpha-2)(3\alpha-2)+2\scut\left(2\alpha(\alpha-2)(11\alpha-14)+8\scut\left(3\alpha(\alpha-2)+\scut\alpha\right)\right)}{\alpha^5}\:.
\end{align}
\hrule
\end{figure*}

Note that the results for calculating $\expect[I^k\exp(-I)]$ for $\indiii\in\n$ for Rayleigh fading can be generalized to any fading distribution with the property $\expect[\fgain_\indi^\falpha(\coord)]<\infty$ due to the equivalence of the propagation process~\cite{6707855}.
This can be done by replacing $\dens$ by $\dens'=\dens\frac{\expect[\fgain_\indi^\falpha(\coord)]}{\Gamma(\falpha+1)}$ in~\eqref{eq:i_expi}, \eqref{eq:i_expi_al_iv}, and \eqref{eq:i2exp}-\eqref{eq:i4exp}.
For example, in the case of Nakagami fading, where $\fgain_\indi(\coord)$ follows a Gamma distribution (see Table~\ref{tab:fading}), the corresponding moment is given by
\begin{equation}
\expect[\fgain_\indi^\falpha(\coord)]=\nakagamim^\falpha\,\frac{\Gamma(\falpha+m)}{\Gamma(\nakagamim)}\:.
\end{equation}

\subsection{Derivations using the Laplace Transform}
%\comment{TODO: move to right place or delete this section.}
%\comment{TODO: use same notation as in rest of paper.}

An alternative approach for deriving the expected value $\expect\big[\interf^\indiii\exp(-s\interf)\big]$, for $\indiii\in\n$ and $s\in\re^+$, which is a special case of the results derived in the previous sections, is by applying the Laplace transform of the interference (cf.~\cite{haenggi09:interference}). We have
\begin{equation}\label{eq:haenggi}
\expect_{\ppp,\fgain,\indicator}\big[\interf^\indiii\exp(-s\interf)\big]=(-1)^\indiii L_\interf^{(\indiii)}(s)\:,
\end{equation}
where $L_\interf(s)$ is the Laplace transform of the interference.

As an example, we consider the Laplace transform for the singular path-loss model and Rayleigh fading (see \cite{haenggi09:interference}, (3.21)) with transmitter density $\txp\dens$ given by
\begin{equation}
L_\interf(s)=\exp\left(-\txp\dens\pi s^\falpha\frac{\pi\falpha}{\sin(\pi\falpha)}\right)\:,
\end{equation}
with $\falpha=2/\alpha$. 
When taking the first derivative of this expression and evaluate $-L_\interf'(1)$, this yields~\eqref{eq:i_expi}.
For the second, third and fourth derivative we get~\eqref{eq:i2exp}-\eqref{eq:i4exp}.

%An interesting question, which we might want to further pursue, is whether there is a similar equation as~\eqref{eq:haenggi} for the $m$-dimensional case $\expect\big[I_1^{n_1}\dots I_m^{n_m}\exp(-s(I_1+\dots+I_m))\big]$ based on the partial derivatives of the $m$-dimensional Laplacian.

\section{Temporal Dependence of Outage under Nakagami Fading}\label{sec:nakagami}
The temporal correlation of link outages under Rayleigh fading has been derived in~\cite{haenggi13:div-poly}.
In the following we derive the result for the more general Nakagami fading model.
For simplification we assume $\nakagamim\in\n$ throughout this section.

\subsection{Derivation of Outage Probabilities}
In this section we apply the following network model:
A source $\src$ transmits data packets to a destination $\dst$ within a stationary Poisson field $\ppp$ of interferers with intensity $\dens$. Let $\dist=\|\dst-\src\|_2$ denote the distance between $\src$ and $\dst$.
Slotted ALOHA is employed, i.e., each interferer transmits in each slot with probability $\txp$.
Fading is assumed to be Nakagami with parameter $\nakagamim$, i.e., $\fgain\sim\Gamma(\nakagamim,\frac{1}{\nakagamim})$ with $\nakagamim>0$.
Let $\sevent_\indiii$ denote the event that a transmission from $\src$ to $\dst$ is successful at slot $\indiii$.
We assume that the event $\sevent_\indiii$ occurs iff $\sir\geq\thr$ for some constant threshold $\thr$, where $\sir=\frac{\fgain_\indiii\gain(\dist)}{\interf_\indiii}$ denotes the signal-to-interference ratio. Here, $\interf_\indiii$ is defined as in~\eqref{eq:interf}.
Further, let $\thri=\frac{\thr}{\gain(\dist)}$ denote the receiver threshold divided by the path gain from $\src$ to $\dst$.
We start by deriving an expression for $\prob[\sevent_\indiii]$.

\begin{theorem}[Transmission success probability]\label{th:nakagami1}\it
The success probability of a single transmission assuming Nakagami fading is
\begin{eqnarray}\label{eq:nakagamii}
\prob[\sevent_\indiii]
%&=&\prob[\fgain_\indiii>\thri \interf_\indiii]
%\\\nonumber
&=&\exp\left(-\txp\dens\int_{\re^2}\Big(1-\big(1+\thri\gain(\coord)\big)^{-\nakagamim}\Big)\,\dd \coord\right)
\\\nonumber&&
\left(1+
\sum_{\indi=1}^{\nakagamim-1}\frac{1}{\indi!}
\sum_{\indlen=1}^{\indi}
\sum_{\mat\in\matseti}
\frac{\magic_\mat}{\indlen!}
\prod_{\indii=1}^\indlen 
\txp\dens\,
\frac{\Gamma(\nakagamim+\melem_{\indii1})}{\Gamma(\nakagamim)}
\int_{\re^2}
\frac{\big(\thri\gain(\coord)\big)^{\melem_{\indii1}}}{\big(1+\thri\gain(\coord)\big)^{\nakagamim+\melem_{\indii1}}}
\,\dd \coord
\right)\:,
\end{eqnarray}
where $\mat\in\n^\indlen$ is a vector of length $\indlen$.
\end{theorem}
\begin{IEEEproof}
The probability of the event $\sevent_\indiii$ in an arbitrary time slot $\indiii$ is given by
\begin{eqnarray}
\prob[\sevent_\indiii]&=&\prob[\fgain_\indiii>\thri \interf_\indiii]\\\nonumber
&\stackrel{(a)}{=}&\expect_{\ppp,\fgain,\indicator}\left[\sum_{\indi=0}^{\nakagamim-1}\frac{1}{\indi!}\left(\nakagamim\thri\interf_\indiii\right)^\indi\exp\left(-\nakagamim\thri\interf_\indiii\right)\right]\\\nonumber
&=&\sum_{\indi=0}^{\nakagamim-1}\frac{1}{\indi!}\left(\nakagamim\thri\right)^\indi\expect_{\ppp,\fgain,\indicator}\left[\interf_\indiii^\indi\exp\left(-\nakagamim\thri\interf_\indiii\right)\right]\\\nonumber
&\stackrel{(b)}{=}&
\exp\left(-\txp\dens\int_{\re^2}\Big(1-\big(1+\thri\gain(\coord)\big)^{-\nakagamim}\Big)\,\dd \coord\right)
\\\nonumber&&
\left(1+
\sum_{\indi=1}^{\nakagamim-1}\frac{1}{\indi!}
\sum_{\indlen=1}^{\indi}
\sum_{\mat\in\matseti}
\frac{\magic_\mat}{\indlen!}
\prod_{\indii=1}^\indlen 
\txp\dens\,
\frac{\Gamma(\nakagamim+\melem_{\indii1})}{\Gamma(\nakagamim)}
\int_{\re^2}
\frac{\big(\thri\gain(\coord)\big)^{\melem_{\indii1}}}{\big(1+\thri\gain(\coord)\big)^{\nakagamim+\melem_{\indii1}}}
\,\dd \coord
\right)\:,
\end{eqnarray}
where in $(a)$ the sum representation of the ccdf of the gamma distribution for integer $\nakagamim$ is substituted. $(b)$ holds due to Theorem~\ref{th:expectmain} for $\indi>0$ and~\eqref{eq:pgfl} for $\indi=0$. Further, we calculate the expected values $\expect_\fgain$ (see Table~\ref{tab:fadingexpect}, Nakagami fading).
\end{IEEEproof}

Note that in this section the letter $m$ is used in two ways: we use $\nakagamim$ as the parameter for fading while $\melem_{\indi\indii}$ indicates an element in the matrix $\mat$.

Next, we derive the joint probability of success in two different time slots $\indiv,\indv$.
\begin{theorem}[Joint transmission success probability]\label{th:nakagami2}\it
Let $\indiv,\indv\in\n$ with $\indiv\neq\indv$ denote two time slots. Then the probability of transmission success in both slots is given by
\begin{eqnarray}\label{eq:nak:joint}
\prob[\sevent_\indiv,\sevent_\indv]&=&
\exp\left(-\dens\int_{\re^2}1-
\left(\txp\big(1+\thri\gain(\coord)\big)^{-
\nakagamim}+1-\txp\right)^2\,\dd \coord\right)
\vast(1+\sum_{\indi=0}^{\nakagamim-1}\sum_{\indii=0\atop\indi+\indii>0}^{\nakagamim-1}\frac{1}{\indi!\indii!}
\\\nonumber&&
\sum_{\indlen=1}^{\maxlenhomo}
\sum_{\mat\in\matset}
\frac{\magic_\mat}{\indlen!}
\prod_{\indiii=1}^\indlen\dens\,
\int_{\re^2}
\left(\txp\,
\frac{\Gamma(\nakagamim+\melem_{\indiii1})\big(\thri\gain(\coord)\big)^{\melem_{\indiii1}}}{\Gamma(\nakagamim)\big(1+\thri\gain(\coord)\big)^{\nakagamim+\melem_{\indiii1}}}
%\frac{\nakagamim^{2\melem_{\indiii1}}\fadOmega^{2\nakagamim}\Gamma(\nakagamim+\melem_{\indiii1})\big(\gain(\coord)\thri\big)^{\melem_{\indiii1}}}{\Gamma(\nakagamim)\big(\fadOmega^2+\gain(\coord)\nakagamim^2\thri\big)^{\nakagamim+\melem_{\indiii1}}}
+(1-\txp)\,\indimi\right)
\\\nonumber&&
\left(\txp\,
\frac{\Gamma(\nakagamim+\melem_{\indiii2})\big(\thri\gain(\coord)\big)^{\melem_{\indiii2}}}{\Gamma(\nakagamim)\big(1+\thri\gain(\coord)\big)^{\nakagamim+\melem_{\indiii2}}}
%\frac{\nakagamim^{2\melem_{\indiii2}}\fadOmega^{2\nakagamim}\Gamma(\nakagamim+\melem_{\indiii2})\big(\gain(\coord)\thri\big)^{\melem_{\indiii2}}}{\Gamma(\nakagamim)\big(\fadOmega^2+\gain(\coord)\nakagamim^2\thri\big)^{\nakagamim+\melem_{\indiii2}}}
+(1-\txp)\,\indimii\right)
\,\dd \coord\vast)
\end{eqnarray}
with symbols defined as in Theorem~\ref{th:expectmain}.
Here, $\indimi$ denotes the indicator variable being one for $\melem_{\indiii1}=0$, and else zero.
\end{theorem}
\begin{IEEEproof}
We start the derivation by substituting the definition of the events $\sevent_\indiv$ and $\sevent_\indv$ and get
\begin{eqnarray}
\prob[\sevent_\indiv,\sevent_\indv]
&=&\prob[\fgain_\indiv>\thri\interf_\indiv,\fgain_\indv>\thri\interf_\indv]\\\nonumber
&=&\expect_{\ppp,\fgain,\indicator}\left[\sum_{\indi=0}^{\nakagamim-1}\frac{1}{\indi!}\left(\nakagamim\thri \interf_\indiv\right)^\indi\exp\left(-\nakagamim\thri \interf_\indiv\right)
%\\\nonumber&&\cdot\,
\sum_{\indii=0}^{\nakagamim-1}\frac{1}{\indii!}\left(\nakagamim\thri\interf_\indv\right)^\indii\exp\left(-\nakagamim\thri\interf_\indv\right)\right]\\\nonumber
&=&\sum_{\indi=0}^{\nakagamim-1}\sum_{\indii=0}^{\nakagamim-1}\frac{1}{\indi!\indii!}\left(\nakagamim\thri\right)^{\indi+\indii} \expect_{\ppp,\fgain,\indicator}\left[\interf_\indiv^\indi\interf_\indv^\indii\exp\left(-\nakagamim\thri (\interf_\indiv+\interf_\indv)\right)\right]\:.
\end{eqnarray}
To be able to calculate this probability we have to derive an expression for the expected value in the last line.
Here we distinguish two cases:
For the case $\indi+\indii>0$ we apply Theorem~\ref{th:expectmain}, which gives
% Let $\pppk$ and $\pppl$ denote the set of active nodes in slot $\nakagamim$ and $l$, respectively. Then, we have
\begin{eqnarray}%\label{eq:nakagamiii}
\lefteqn{\expect_{\ppp,\fgain,\indicator} \left[\interf_\indiv^\indi\interf_\indv^
\indii\exp\left(-\nakagamim\thri (\interf_\indiv+\interf_
\indv)\right)\right]}\\\nonumber
&=&\expect_{\ppp,\fgain,\indicator}\Bigg[\Bigg(\sum_{\pointi\in\ppp}\fgain_\indiv(\pointi
)\gain(\pointi)\indiciv\Bigg)^\indi\Bigg(\sum_{\pointii\in\ppp}\fgain_\indv(\pointii)\gain
(\pointii)\indicv\Bigg)^\indii 
\\\nonumber&&
\exp\left(-\nakagamim\thri\sum_{
\pointiii\in\ppp}\gain(\indiii)\big(\fgain_\indiv(\pointiii)\indicwiv+\fgain_\indv(
\pointiii)\indicwv\big)\right)\Bigg]\\\nonumber
&\stackrel{(a)}{=}&\exp\Bigg(-\dens\int_{\re^2}\bigg(1-
\left(\txp\,\expect_{\fgain_\indiv(\coord)}\left[\exp\left(-\nakagamim\thri\fgain_\indiv(\coord)\gain(\coord)\right)\right]+1-\txp\right)
\\\nonumber&&
\left(\txp\,\expect_{\fgain_\indv(\coord)}\left[\exp\left(-\nakagamim\thri\fgain_\indv(\coord)\gain(\coord)\right)\right]+1-\txp\right)
\bigg)\,\dd \coord\Bigg)
\\\nonumber&&
\sum_{\indlen=1}^{\maxlenhomo}
\sum_{\mat\in\matset}
\frac{\magic_\mat}{\indlen!}
\prod_{\indiii=1}^\indlen \dens\int_{\re^2}
\left(\txp\,\expect_{\fgain_\indiv(\coord)}\left[\exp\left(-\nakagamim\thri\fgain_\indiv(\coord)\gain(\coord)\right)\big(\fgain_\indiv(\coord)
\gain(\coord)\big)^{\melem_{\indiii1}}\right]+(1-\txp)\,\indimi\right)
\\\nonumber&&
\left(\txp\,\expect_{\fgain_\indv(\coord)}\left[\exp\left(-\nakagamim\thri\fgain_\indv(\coord)\gain(\coord)\right)\big(\fgain_\indv(\coord)
\gain(\coord)\big)^{\melem_{\indiii2}}\right]+(1-\txp)\,\indimii\right)
\,\dd \coord\\\nonumber
&\stackrel{(b)}{=}&\exp\left(-\dens\int_{\re^2}1-
\left(\txp\big(1+\thri\gain(\coord)\big)^{-
\nakagamim}+1-\txp\right)^2\,\dd \coord\right)
\\\nonumber&&
\sum_{\indlen=1}^{\maxlenhomo}
\sum_{\mat\in\matset}
\frac{\magic_\mat}{\indlen!}
\prod_{\indiii=1}^\indlen\dens\,
\int_{\re^2}
\left(\txp\,
\frac{\Gamma(\nakagamim+\melem_{\indiii1})\big(\thri\gain(\coord)\big)^{\melem_{\indiii1}}}{\Gamma(\nakagamim)\big(1+\thri\gain(\coord)\big)^{\nakagamim+\melem_{\indiii1}}}
%\frac{\nakagamim^{2\melem_{\indiii1}}\fadOmega^{2\nakagamim}\Gamma(\nakagamim+\melem_{\indiii1})\big(\gain(\coord)\thri\big)^{\melem_{\indiii1}}}{\Gamma(\nakagamim)\big(\fadOmega^2+\gain(\coord)\nakagamim^2\thri\big)^{\nakagamim+\melem_{\indiii1}}}
+(1-\txp)\,\indimi\right)
\\\nonumber&&
\left(\txp\,
\frac{\Gamma(\nakagamim+\melem_{\indiii2})\big(\thri\gain(\coord)\big)^{\melem_{\indiii2}}}{\Gamma(\nakagamim)\big(1+\thri\gain(\coord)\big)^{\nakagamim+\melem_{\indiii2}}}
%\frac{\nakagamim^{2\melem_{\indiii2}}\fadOmega^{2\nakagamim}\Gamma(\nakagamim+\melem_{\indiii2})\big(\gain(\coord)\thri\big)^{\melem_{\indiii2}}}{\Gamma(\nakagamim)\big(\fadOmega^2+\gain(\coord)\nakagamim^2\thri\big)^{\nakagamim+\melem_{\indiii2}}}
+(1-\txp)\,\indimii\right)
\,\dd \coord\:,
\end{eqnarray}
where $\fexp=(\indi,\indii)$.
In the above expression, $(a)$ holds due to Theorem~\ref{th:expectmain} and in $(b)$ we calculated the expected values of the gamma distributed random variables $\fgain\sim\Gamma\left(\nakagamim,\frac{1}{\nakagamim}\right)$.

For the case $\indi+\indii=0$ we apply~\eqref{eq:pgfl}, such that
\begin{eqnarray}\label{eq:nakagamiiii}
\lefteqn{\expect_{\ppp,\fgain,\indicator} \left[\exp\left(-\nakagamim\thri (\interf_\indiv+\interf_
\indv)\right)\right]}\\\nonumber
&\stackrel{(a)}{=}&\exp\vast(-\dens\int_{\re^2}\Bigg(1-
\left(\txp\,\expect_{\fgain_\indiv(\coord)}\left[\exp\left(-\nakagamim\thri\fgain_\indiv(\coord)\gain(\coord)\right)\right]+1-\txp\right)
\\\nonumber&&
\left(\txp\,\expect_{\fgain_\indv(\coord)}\left[\exp\left(-\nakagamim\thri\fgain_\indv(\coord)\gain(\coord)\right)\right]+1-\txp\right)
\Bigg)\,\dd \coord\vast)
\\\nonumber
&\stackrel{(b)}{=}&\exp\left(-\dens\int_{\re^2}1-
\left(\txp\big(1+\thri\gain(\coord)\big)^{-
\nakagamim}+1-\txp\right)^2\,\dd \coord\right)\:.
\end{eqnarray}
Here, $(a)$ holds due to~\eqref{eq:pgfl} and in $(b)$ we calculate the expected values over $\fgain\sim\Gamma\left(\nakagamim,\frac{1}{\nakagamim}\right)$.
\end{IEEEproof}

If we, for example, substitute the singular path-loss model into~\eqref{eq:nakagamii}, we get the following result.
\begin{corollary}[Transmission success probability with singular path loss]\it
For the singular path loss model $\gain(\coord)=\|\coord\|_2^{-\plexp}$ we have
\begin{eqnarray}\label{eq:nakagami:sing}
\prob[\sevent_\indiii]&=&
\exp\left(-\txp\dens\,
\frac{\pi\thri^{\falpha}\Gamma(1-\falpha)\Gamma(\nakagamim+\falpha)}{\Gamma(\nakagamim)}
\right)
\\\nonumber&&
\left(1+\sum_{\indi=1}^{\nakagamim-1}\frac{1}{\indi!}
\sum_{\indlen=1}^{\indi}
\left(\frac{\txp\dens\falpha\pi \thri^\falpha\Gamma(\nakagamim+\falpha)}{\Gamma(\nakagamim)}\right)^\indlen
\sum_{\mat\in\matseti}
\frac{\magic_\mat}{\indlen!}
\prod_{\indii=1}^\indlen 
\Gamma(\melem_{\indii1}-\falpha)
%
%\frac{\left(\frac{\fadOmega^2}{\nakagamim^2\thri}\right)^{\melem_{\indii1}-\falpha}\Gamma(\melem_{\indii1}-\falpha)\Gamma(\nakagamim+\falpha)(\nakagamim^2\thri)^{\melem_{\indii1}}\fadOmega^{-2\melem_{\indii1}}}{\plexp\Gamma(\nakagamim)}
%
%
\right)\:.
\end{eqnarray}
\end{corollary}
\begin{IEEEproof}
Substituting $\gain(\coord)=\|\coord\|_2^{-\plexp}$ and $\txp=1$ into Theorem~\ref{th:nakagami1} yields the result.
\end{IEEEproof}

Next, we substitute the singular path-loss model and $\txp=1$ into~\eqref{eq:nak:joint}.
\begin{corollary}[Joint transmission success probability with singular path loss]\it
For the singular path loss model $\gain(\coord)=\|\coord\|_2^{-\plexp}$ and $\txp=1$ we have
\begin{eqnarray}\label{eq:nakagami:sing2}
\prob[\sevent_\indiv,\sevent_\indv]
&=&
\exp\left(-\dens
\frac{\pi\thri^{\falpha}\Gamma(1-\falpha)\Gamma(2\nakagamim+\falpha)}{\Gamma(2\nakagamim)}
\right)
\vast(1+\sum_{\indi=0}^{\nakagamim-1}\sum_{\indii=0\atop\indi+\indii>0}^{\nakagamim-1}\frac{1}{\indi!\indii!}
%\\\nonumber&&
\sum_{\indlen=1}^{\indi+\indii}
\left(\frac{\dens\falpha\pi\thri^\falpha\Gamma(2m+\falpha)}{\Gamma^2(\nakagamim)}\right)^\indlen
\\\nonumber&&
\sum_{\mat\in\matsetii}
\frac{\magic_\mat}{\indlen!}
\prod_{\indiii=1}^\indlen
\frac{\Gamma(\nakagamim+\melem_{\indiii1})\Gamma(\nakagamim+\melem_{\indiii2})\Gamma(\melem_{\indiii1}+\melem_{\indiii2}-\falpha)}{\Gamma(\melem_{\indiii1}+\melem_{\indiii2}+2\nakagamim)}
\vast)\:.
\end{eqnarray}
\end{corollary}
\begin{IEEEproof}
Substituting $\gain(\coord)=\|\coord\|_2^{-\plexp}$ and $\txp=1$ into Theorem~\ref{th:nakagami2} yields the result.
\end{IEEEproof}

\subsection{Outage Probability for the Singular Path Loss}

\begin{figure}[b!]
\centering
\pgfplotsset{scaled x ticks=false}
\begin{tikzpicture}[scale=1]
\begin{semilogyaxis}[xlabel={Intensity $\dens$},
	ylabel={Outage probability $\prob[\bar{\sevent}_\indiii]$},ymin=0.001,ymax=1,xmin=0,xmax=0.03,grid=both,xtick={0.005,0.01,0.015,0.02,0.025,0.03},
            ytick={0.0001,0.001,0.01,0.1,1},
xticklabels={0.005,0.01,0.015,0.02,0.025,0.03},
yticklabels={$10^{-4}$,$10^{-3}$,$10^{-2}$,$10^{-1}$,$1$},
legend style={at={(0.98,0.02)}, anchor=south east, font=\footnotesize},
legend cell align=left,
legend entries={$\alpha=2.5$,
                $\alpha=3$,
                $\alpha=4$,
                $\alpha=5$}]
								
\addlegendimage{solid,style=thick,mark=x,mark options={thick,solid}}
\addlegendimage{dashed,style=thick,mark=diamond,mark options={thick,solid}}
\addlegendimage{densely dotted,style=thick, mark=triangle,mark options={thick,solid}}
\addlegendimage{dashdotted,style=thick,mark=square,mark options={thick,solid}}

\addplot plot[color=black,solid,no marks,style=thick]  table[x index=0,y index=1]  {P_outage1_Nakagami.txt};
\addplot plot[color=black,only marks,mark=x,mark options={solid}] table[x index=0,y index=3] {fig-2-alpha-2dot5.dat};
\addplot plot[color=black,dashed,no marks,style=thick]  table[x index=0,y index=2]  {P_outage1_Nakagami.txt};
\addplot plot[color=black,only marks,mark=diamond,mark options={solid}] table[x index=0,y index=3] {fig-2-alpha-3.dat};
%\addplot plot[color=black,dotted,no marks,style=thick]  table[x index=0,y index=3]  {P_outage1_Nakagami.txt};\addlegendentry{ ~$\alpha=3.5$ }
\addplot plot[color=black,densely dotted,no marks,style=thick]  table[x index=0,y index=4]  {P_outage1_Nakagami.txt};
\addplot plot[color=black,only marks,mark=triangle,mark options={solid}] table[x index=0,y index=3] {fig-2-alpha-4.dat};
%\addplot plot[color=black,densely dashed,no marks,style=thick]  table[x index=0,y index=5]  {P_outage1_Nakagami.txt};\addlegendentry{ ~$\alpha=4.5$ }
\addplot plot[color=black,dashdotted,no marks,style=thick]  table[x index=0,y index=6]  {P_outage1_Nakagami.txt};
\addplot plot[color=black,only marks,mark=square,mark options={solid}] table[x index=0,y index=3] {fig-2-alpha-5.dat};

\end{semilogyaxis}
\end{tikzpicture}
\caption{Outage probability $\prob[\bar{\sevent}_\indiii]$ given in~\eqref{eq:nakagami:sing} over the interferer intensity $\dens$. Lines indicate the theoretical results while marks are simulation results. Parameters are $\nakagamim=3$, $\thr=0.5$, and $\dist=2$.}\label{fig:Psucc:Nakagami:dens}
\end{figure}

\begin{figure}[b!]
\centering
\begin{tikzpicture}[scale=1]
\begin{axis}[xlabel={Path loss exponent $\plexp$},
	ylabel={Success probability $\prob[\sevent_\indiii]$},ymin=0,ymax=0.7,xmin=2,xmax=4,grid=both,xtick={2,2.5,3,3.5,4,4.5,5},
            ytick={0,0.1,0.2,0.3,0.4,0.5,0.6,0.7},
xticklabels={2,2.5,3,3.5,4,4.5,5},
yticklabels={0,0.1,0.2,0.3,0.4,0.5,0.6,0.7},
legend style={at={(0.98,0.02)}, anchor=south east, font=\footnotesize},
legend cell align=left
]

\addplot plot[color=black,solid,no marks,style=thick]  table[x index=0,y index=1]  {P_succ_Nakagami_alpha.txt};\addlegendentry{ ~$\nakagamim=1$ }
\addplot plot[color=black,dotted,no marks,style=thick]  table[x index=0,y index=2]  {P_succ_Nakagami_alpha.txt};\addlegendentry{ ~$\nakagamim=2$ }
\addplot plot[color=black,densely dotted,no marks,style=thick]  table[x index=0,y index=3]  {P_succ_Nakagami_alpha.txt};\addlegendentry{ ~$\nakagamim=3$ }
\addplot plot[color=black,densely dashed,no marks,style=thick]  table[x index=0,y index=4]  {P_succ_Nakagami_alpha.txt};\addlegendentry{ ~$\nakagamim=4$ }
\addplot plot[color=black,dashdotted,no marks,style=thick]  table[x index=0,y index=5]  {P_succ_Nakagami_alpha.txt};\addlegendentry{ ~$\nakagamim=5$ }

\end{axis}
\end{tikzpicture}
\caption{Transmission success probability $\prob[\sevent_\indiii]$ given in~\eqref{eq:nakagami:sing} over the path loss exponent $\plexp$. Parameters are $\dens=0.03$, $\thr=1$, and $\dist=2$.}\label{fig:Psucc:Nakagami:alpha}
\end{figure}

In the following, plots of \eqref{eq:nakagami:sing} and~\eqref{eq:nakagami:sing2} are presented. Fig.~\ref{fig:Psucc:Nakagami:dens} shows the outage probability $\prob[\bar{\sevent}_\indiii]=1-\prob[\sevent_\indiii]$ over the interferers' intensity $\dens$. The following observations can be made:
Firstly, the outage probability is higher for higher intensity $\dens$ and hence higher interference, with a non-linear dependence.
In the limit, the outage probability approaches one, i.e., $\lim_{\dens\to\infty}(1-\prob[\sevent_\indiii])=1$.
Secondly, for high values of the path loss exponent $\plexp$ the attenuation of both the received signal and the interference is higher than for low values. In the scenario presented in Fig.~\ref{fig:Psucc:Nakagami:dens}, the attenuation effect is stronger for interference than for the received signal. Hence, the outage probability is lower for high path loss exponents $\plexp$.

In addition to the theoretical results, Fig.~\ref{fig:Psucc:Nakagami:dens} also shows simulation results to provide a validation of the results. As can be seen, simulations and theory do indeed match very well.

\begin{figure}[t!]
\centering
\pgfplotsset{scaled x ticks=false}
\begin{tikzpicture}[scale=1]
\begin{semilogyaxis}[xlabel={Intensity $\dens$},
	ylabel={Outage probability $\prob[\bar{\sevent}_\indiii]$},ymin=0.001,ymax=1,xmin=0,xmax=0.1,grid=both,xtick={0.02,0.04,0.06,0.08,0.1},
            ytick={0.0001,0.001,0.01,0.1,1},
xticklabels={0.02,0.04,0.06,0.08,0.1},
yticklabels={$10^{-4}$,$10^{-3}$,$10^{-2}$,$10^{-1}$,$1$},
legend style={at={(0.98,0.02)}, anchor=south east, font=\footnotesize},
legend cell align=left,
legend entries={$\nakagamim=1$,
                $\nakagamim=2$,
                $\nakagamim=3$,
                $\nakagamim=4$,
								$\nakagamim=5$}]
								
\addlegendimage{solid,style=thick,no marks}
\addlegendimage{dashed,style=thick,no marks}
\addlegendimage{densely dotted,style=thick,no marks}
\addlegendimage{dashdotted,style=thick,no marks}
\addlegendimage{dash pattern=on 2pt off 2pt,style=thick,no marks}

\addplot plot[color=black,solid,no marks,style=thick]  table[x index=0,y index=1]  {P_Rev3_outage_nakagami_m.txt};
\addplot plot[color=black,dashed,no marks,style=thick]  table[x index=0,y index=2]  {P_Rev3_outage_nakagami_m.txt};
\addplot plot[color=black,densely dotted,no marks,style=thick]  table[x index=0,y index=3]  {P_Rev3_outage_nakagami_m.txt};
\addplot plot[color=black,dashdotted,no marks,style=thick]  table[x index=0,y index=4]  {P_Rev3_outage_nakagami_m.txt};
\addplot plot[color=black,dash pattern=on 2pt off 2pt,no marks,style=thick]  table[x index=0,y index=5]  {P_Rev3_outage_nakagami_m.txt};

\end{semilogyaxis}
\end{tikzpicture}
\caption{Outage probability $\prob[\bar{\sevent}_\indiii]$ given in~\eqref{eq:nakagami:sing} over the interferer intensity $\dens$. Parameters are $\plexp=3$, $\thr=0.5$, and $\dist=2$.}\label{fig:Psucc:Nakagami:m}
\end{figure}

\begin{figure}[t!]
\centering
\pgfplotsset{scaled x ticks=false}
\begin{tikzpicture}[scale=1]
\begin{semilogyaxis}[xlabel={Path loss exponent $\plexp$},
	ylabel={Outage probability $\prob[\bar{\sevent}_\indiii]$},ymin=0.01,ymax=1,xmin=2,xmax=5,grid=both,xtick={2,3,4,5},
            ytick={0.0001,0.001,0.01,0.1,1},
xticklabels={2,3,4,5},
yticklabels={$10^{-4}$,$10^{-3}$,$10^{-2}$,$10^{-1}$,$1$},
legend style={at={(0.98,0.98)}, anchor=north east, font=\footnotesize},
legend cell align=left,
legend entries={$\nakagamim=1$,
                $\nakagamim=2$,
                $\nakagamim=3$,
                $\nakagamim=4$,
								$\nakagamim=5$}]
								
\addlegendimage{solid,style=thick,no marks}
\addlegendimage{dashed,style=thick,no marks}
\addlegendimage{densely dotted,style=thick,no marks}
\addlegendimage{dashdotted,style=thick,no marks}
\addlegendimage{dash pattern=on 2pt off 2pt,style=thick,no marks}

\addplot plot[color=black,solid,no marks,style=thick]  table[x index=0,y index=1]  {P_Rev3_outage_nakagami_strange.txt};
\addplot plot[color=black,dashed,no marks,style=thick]  table[x index=0,y index=2]  {P_Rev3_outage_nakagami_strange.txt};
\addplot plot[color=black,densely dotted,no marks,style=thick]  table[x index=0,y index=3]  {P_Rev3_outage_nakagami_strange.txt};
\addplot plot[color=black,dashdotted,no marks,style=thick]  table[x index=0,y index=4]  {P_Rev3_outage_nakagami_strange.txt};
\addplot plot[color=black,dash pattern=on 2pt off 2pt,no marks,style=thick]  table[x index=0,y index=5]  {P_Rev3_outage_nakagami_strange.txt};

\end{semilogyaxis}
\end{tikzpicture}
\caption{Outage probability $\prob[\bar{\sevent}_\indiii]$ given in~\eqref{eq:nakagami:sing} over the path loss exponent $\plexp$. Parameters are $\dens=0.01$, $\thr=0.5$, and $\dist=\sqrt[\plexp]{4}$.}\label{fig:Psucc:Nakagami:strange}
\end{figure}

To explore this effect in more detail, in Fig.~\ref{fig:Psucc:Nakagami:alpha} the impact of the path loss exponent $\plexp$ is shown. As can be seen, the dependence of $\prob[\sevent_\indiii]$ on $\plexp$ is monotonic for the plotted parameters. For $\plexp$ close to two the success probability is very low and approaching zero, as expected in case of a stationary PPP of interferers on the plane; for higher values it is monotonically increasing. If we further increase $\plexp$ beyond the range of the plot, the success probability approaches a limiting value, which is given by $\lim_{\plexp\to\infty}\prob[\sevent_\indiii]=\exp(-\mu)$ with $\mu=\dist^2\pi\dens$ being the expected number of interferers in a circle with radius $\dist$ around $\dst$. This behavior can be explained by recalling the modeling assumptions: A successful transmission is defined as the event that the SIR is above a certain threshold, i.e., neither consider noise nor receiver sensitivity.
As the path loss gets large, the success probability exhibits a hard-core behavior, since any node closer to the receiver than the transmitter would cause overwhelming interference. Hence, the success probability is equal to the probability that there is no interferer in the disk of radius $\dist$ centered at the destination.
%Hence, although for $\plexp\to\infty$ both the received signal and the interference go to zero, we still get a strictly positive success probability. 
%When considering noise or a receiver sensitivity, the limiting value would change to zero.

Another interesting observation can be made in Fig.~\ref{fig:Psucc:Nakagami:alpha}: The two parameters $\plexp$ and $\nakagamim$ have a joint impact on the success probability. For small $\plexp$, more severe fading, i.e., smaller values of $\nakagamim$, leads to high success probabilities. This can be explained by the fact that small $\plexp$ lead to strong interference, which can be partly mitigated by harsh fading conditions. Hence, in this regime fading diminishes interference stronger than it diminishes the received signal. For higher values of $\plexp$ this trend is inverted. Here, although interference can still be reduced by severe fading, its impact on the received signal is dominant. In between there is a certain value of $\plexp$ for which the parameter $\nakagamim$ plays no role at all. This value depends on the parameters $\dens$, $\thr$, and $\dist$.

Further, in Fig.~\ref{fig:Psucc:Nakagami:m} a plot of the outage probability over the interferer intensity $\dens$ for different values of the Nakagami parameter $\nakagamim$ is shown. Again, we see that for a certain value of $\dens$ (which is at about $\dens=0.0742$), similar as for $\alpha$ in Fig.~\ref{fig:Psucc:Nakagami:alpha}, the curves for different $\nakagamim$ intersect at a common point. For densities $\dens$ below this threshold outage increases with $\nakagamim$, while for $\dens$ above this threshold this trend is inverted.

Finally, Fig.~\ref{fig:Psucc:Nakagami:strange} shows a plot of the outage probability over the path loss exponent $\plexp$ for different values of $\nakagamim$. The special feature of this plot is that the distance between transmitter and receiver $\dist$ is chosen in a way such that $\dist^\plexp=4$ over the whole range of values for $\plexp$.
Again, the lines overlap at a common point, which is at about $\plexp=2.1$. When further increasing $\plexp$, the difference of the outage probabilities for different $\nakagamim$ increases to a maximum, and then starts decreasing.
When further increasing the path loss exponent $\plexp$, with the given parameters the outage probability approaches the limit $\lim_{\plexp\to\infty}\prob[\bar{\sevent}_\indiii]\approx 0.0309276$, independent of $\nakagamim$.

\subsection{Joint Outage Probability for the Singular Path Loss}
Next, we investigate the joint outage probability of two transmissions in different time slots. Let therefore $\bar{\sevent}_\indiv$ denote the complementary event of $\sevent_\indiv$, i.e., that the transmission in slot $\indiv$ is in outage. We can calculate the joint outage probability by $\prob[\bar{\sevent}_\indiv,\bar{\sevent}_\indv]=1-\prob[\sevent_\indiv]-\prob[\sevent_\indv]+\prob[\sevent_\indiv,\sevent_\indv]$, which is shown in Fig.~\ref{fig:Psucc:Nakagami:dens2}.
Overall, the plot shows similar trends as the one in Fig.~\ref{fig:Psucc:Nakagami:dens}, with the obvious difference that for the given parameters the joint outage probability $\prob[\bar{\sevent}_\indiv,\bar{\sevent}_\indv]$ is smaller than the outage probability $1-\prob[\sevent_\indiii]$ of a single transmission for all $\dens>0$. 

There is one small detail, which is very interesting in Fig.~\ref{fig:Psucc:Nakagami:dens2}: Similar to the impact of the fading parameter $\nakagamim$ (see Fig.~\ref{fig:Psucc:Nakagami:alpha}), also the influence of the path loss exponent $\plexp$ on the outage probability is determined by the values of other parameters. In particular, for small intensities $\dens$ --- in the low interference regime --- lower path loss exponents are beneficial (left side of the plot). For high intensities $\dens$ --- in the high interference regime --- this trend is inverted (right side of the plot). Here, high path loss exponents $\plexp$ significantly reduce the interference; and this effect is stronger than the degradation of the received signal due to the higher $\plexp$. Between these two extremes there is a non-monotonic dependence on $\plexp$, as can be seen in the plot, e.g., for $\dens=0.015$.

\begin{figure}[t!]
\centering
\pgfplotsset{scaled x ticks=false}
\begin{tikzpicture}[scale=1]
\begin{semilogyaxis}[xlabel={Intensity $\dens$},
	ylabel={Joint outage probability $\prob[\bar{\sevent}_\indiv,\bar{\sevent}_\indv]$},ymin=0.001,ymax=1,xmin=0,xmax=0.03,grid=both,xtick={0.005,0.01,0.015,0.02,0.025,0.03},
            ytick={0.0001,0.001,0.01,0.1,1},
xticklabels={0.005,0.01,0.015,0.02,0.025,0.03},
yticklabels={$10^{-4}$,$10^{-3}$,$10^{-2}$,$10^{-1}$,$1$},
legend style={at={(0.98,0.02)}, anchor=south east, font=\footnotesize},
legend cell align=left,
legend entries={$\alpha=2.5$,
                $\alpha=3$,
                $\alpha=4$,
                $\alpha=5$}]
								
\addlegendimage{solid,style=thick,mark=x,mark options={thick,solid}}
\addlegendimage{dashed,style=thick,mark=diamond,mark options={thick,solid}}
\addlegendimage{densely dotted,style=thick, mark=triangle,mark options={thick,solid}}
\addlegendimage{dashdotted,style=thick,mark=square,mark options={thick,solid}}

\addplot plot[color=black,solid,no marks,style=thick]  table[x index=0,y index=1]  {P_outage2_Nakagami.txt};\addlegendentry{ ~$\alpha=2.5$ }
\addplot plot[color=black,only marks,mark=x,mark options={solid}] table[x index=0,y index=4] {corr-alpha-2dot5.dat};
\addplot plot[color=black,dashed,no marks,style=thick]  table[x index=0,y index=2]  {P_outage2_Nakagami.txt};\addlegendentry{ ~$\alpha=3$ }
\addplot plot[color=black,only marks,mark=diamont,mark options={solid}] table[x index=0,y index=4] {corr-alpha-3.dat};
\addplot plot[color=black,densely dotted,no marks,style=thick]  table[x index=0,y index=4]  {P_outage2_Nakagami.txt};\addlegendentry{ ~$\alpha=4$ }
\addplot plot[color=black,only marks,mark=triangle,mark options={solid}] table[x index=0,y index=4] {corr-alpha-4.dat};
\addplot plot[color=black,dashdotted,no marks,style=thick]  table[x index=0,y index=6]  {P_outage2_Nakagami.txt};\addlegendentry{ ~$\alpha=5$ }
\addplot plot[color=black,only marks,mark=square,mark options={solid}] table[x index=0,y index=4] {corr-alpha-5.dat};
\end{semilogyaxis}
\end{tikzpicture}
\caption{Probability of two transmissions being in outage $\prob[\bar{\sevent}_\indiv,\bar{\sevent}_\indv]$ given in~\eqref{eq:nakagami:sing2} over interferer intensity $\dens$. Lines indicate theoretical results while marks show simulation results. Parameters are $\nakagamim=3$, $\thr=0.5$, and $\dist=2$.}\label{fig:Psucc:Nakagami:dens2}
\end{figure}

\begin{figure}[t!]
\centering
\begin{tikzpicture}[scale=1]
\begin{axis}[xlabel={Intensity $\dens$},
	ylabel={Success probabilities $(\prob[\sevent_\indiii])^2$, $\prob[\sevent_\indiv,\sevent_\indv]$},ymin=0,ymax=1,xmin=0,xmax=0.1,grid=both,xtick={0,0.02,0.04,0.06,0.08,0.1},
            ytick={0,0.2,0.4,0.6,0.8,1},
xticklabels={0,0.02,0.04,0.06,0.08,0.1},
yticklabels={0,0.2,0.4,0.6,0.8,1},
legend style={at={(0.98,0.98)}, anchor=north east, font=\footnotesize},
legend cell align=left]
								
\addlegendimage{solid,style=thick,mark=x,mark options={thick,solid}}
\addlegendimage{dashed,style=thick,mark=diamond,mark options={thick,solid}}
\addlegendimage{dotted,style=thick, mark=square,mark options={thick,solid}}
\addlegendimage{dashdotted,style=thick,mark=star,mark options={thick,solid}}
\addlegendimage{dash pattern=on 2pt off 2pt,style=thick,mark=triangle,mark options={thick,solid}}
\addlegendimage{densely dotted,style=thick,mark=o,mark options={thick,solid}}

\addplot plot[color=black,solid,no marks,style=thick]  table[x index=0,y index=6]  {P_succ_Nakagami_comp.txt};\addlegendentry{ ~$\prob[\sevent_\indiv,\sevent_\indv]$, $\alpha=5$ }
\addplot plot[color=black,dashed,no marks,style=thick]  table[x index=0,y index=2]  {P_succ_Nakagami_comp.txt};\addlegendentry{ ~$\prob[\sevent_\indiv,\sevent_\indv]$, $\alpha=3$ }
\addplot plot[color=black,dotted,no marks,style=thick]  table[x index=0,y index=1]  {P_succ_Nakagami_comp.txt};\addlegendentry{ ~$\prob[\sevent_\indiv,\sevent_\indv]$, $\alpha=2.5$ }

\addplot plot[color=black,only marks,mark=x,mark options={solid}] table[x index=0,y index=5] {corr-alpha-5.dat};
\addplot plot[color=black,only marks,mark=diamond,mark options={solid}] table[x index=0,y index=5] {corr-alpha-3.dat};
\addplot plot[color=black,only marks,mark=square,mark options={solid}] table[x index=0,y index=5] {corr-alpha-2dot5.dat};

\addplot plot[color=black,dashdotted,no marks, style=thick]  table[x index=0,y index=12]  {P_succ_Nakagami_comp.txt};\addlegendentry{ ~$(\prob[\sevent_\indiii])^2$, $\alpha=5$ }
\addplot plot[color=black,dash pattern=on 2pt off 2pt,no marks, style=thick]  table[x index=0,y index=8]  {P_succ_Nakagami_comp.txt};\addlegendentry{ ~$(\prob[\sevent_\indiii])^2$, $\alpha=3$ }
\addplot plot[color=black,densely dotted,no marks, style=thick]  table[x index=0,y index=7]  {P_succ_Nakagami_comp.txt};\addlegendentry{ ~$(\prob[\sevent_\indiii])^2$, $\alpha=2.5$ }

\addplot plot[color=black,only marks,mark=star,mark options={solid}] table[x index=0,y expr=\thisrowno{3}*\thisrowno{3}] {corr-alpha-5.dat};
\addplot plot[color=black,only marks,mark=triangle,mark options={solid}] table[x index=0,y expr=\thisrowno{3}*\thisrowno{3}] {corr-alpha-3.dat};
\addplot plot[color=black,only marks,mark=o,mark options={solid}] table[x index=0,y expr=\thisrowno{3}*\thisrowno{3}] {corr-alpha-2dot5.dat};

\end{axis}
\end{tikzpicture}
\caption{A comparison between the dependent interference case $\prob[\sevent_\indiv,\sevent_\indv]$ (lines) given in~\eqref{eq:nakagami:sing2} and the independent interference case $(\prob[\sevent_\indiii])^2$ (marks) given in~\eqref{eq:nakagami:sing} over interferer intensity $\dens$. Lines indicate analytical results while marks indicate simulation results. Parameters are $\nakagamim=3$, $\thr=0.5$, and $\dist=2$.}\label{fig:Psucc:Nakagami:dens:comp}
\end{figure}

Next, we compare the probability of two transmissions both being successful for the following two cases: Firstly, interference is dependent due to the same set of interferers. In this case the joint success probability is given by $\prob[\sevent_\indiv,\sevent_\indv]$.
Secondly, interference is assumed to be independent. Here, the joint success probability can be simply calculated by $\big(\prob[\sevent_\indiii]\big)^2$. This case is presented for comparison reasons and to highlight the impact of correlated interference on the success probabilities. It resembles the scenario where interferers are mobile and the time slots a far away from each other.
A plot of these expressions is presented in Fig.~\ref{fig:Psucc:Nakagami:dens:comp}. We can see that for equal parameters the dependent interference case always shows higher values than the independent interference case. This effect stems from the positive correlation of interference in the two time slots $\indiv$ and $\indv$. Similar effects occur in the case of Rayleigh fading and cooperative relaying, as shown in~\cite{schilcher2013coop}.

\begin{figure}[t!]
\centering
\begin{tikzpicture}[scale=1]
\begin{axis}[xlabel={Intensity $\dens$},
	ylabel={Success probability},ymin=0,ymax=1,xmin=0,xmax=0.1,grid=both,xtick={0,0.02,0.04,0.06,0.08,0.1},
            ytick={0,0.2,0.4,0.6,0.8,1},
xticklabels={0,0.02,0.04,0.06,0.08,0.1},
yticklabels={0,0.2,0.4,0.6,0.8,1},
legend style={at={(0.02,0.02)}, anchor=south west, font=\footnotesize},
legend cell align=left
]

\addplot plot[color=black,solid,no marks,style=thick]  table[x index=0,y index=6]  {time_diversity.txt};\addlegendentry{ ~independent , $\alpha=5$ }
%\addplot plot[color=black,dashed,no marks,style=thick]  table[x index=0,y index=5]  {time_diversity.txt};\addlegendentry{ ~independent , $\alpha=4.5$ }
%\addplot plot[color=black,dotted,no marks,style=thick]  table[x index=0,y index=4]  {time_diversity.txt};\addlegendentry{ ~independent , $\alpha=4$ }
%\addplot plot[color=black,densely dotted,no marks,style=thick]  table[x index=0,y index=3]  {time_diversity.txt};\addlegendentry{ ~independent , $\alpha=3.5$ }
\addplot plot[color=black,densely dashed,no marks,style=thick]  table[x index=0,y index=2]  {time_diversity.txt};\addlegendentry{ ~independent , $\alpha=3$ }
\addplot plot[color=black,dashdotted,no marks,style=thick]  table[x index=0,y index=1]  {time_diversity.txt};\addlegendentry{ ~independent , $\alpha=2.5$ }

\addplot plot[color=black,only marks,mark options={solid,scale=0.8}]  table[x index=0,y index=12]  {time_diversity.txt};\addlegendentry{ ~dependent , $\alpha=5$ }
%\addplot plot[color=black,only marks,mark options={scale=0.8}]  table[x index=0,y index=11]  {time_diversity.txt};\addlegendentry{ ~dependent , $\alpha=4.5$ }
%\addplot plot[color=black,only marks,mark options={scale=0.8},mark=+]  table[x index=0,y index=10]  {time_diversity.txt};\addlegendentry{ ~dependent , $\alpha=4$ }
%\addplot plot[color=black,only marks,mark options={scale=0.8}]  table[x index=0,y index=9]  {time_diversity.txt};\addlegendentry{ ~dependent , $\alpha=3.5$ }
\addplot plot[color=black,only marks,mark options={solid,scale=0.8}]  table[x index=0,y index=8]  {time_diversity.txt};\addlegendentry{ ~dependent , $\alpha=3$ }
\addplot plot[color=black,only marks,mark options={solid,scale=0.8},mark=x]  table[x index=0,y index=7]  {time_diversity.txt};\addlegendentry{ ~dependent , $\alpha=2.5$ }

\end{axis}
\end{tikzpicture}
\caption{The probability that at least one out of two transmissions is successfully received over interferer intensity $\dens$. Lines indicate independent interference while marks indicate dependent interference. Parameters are $\nakagamim=3$, $\thr=0.5$, and $\dist=2$.}\label{fig:Psucc:Nakagami:timediv}
\end{figure}
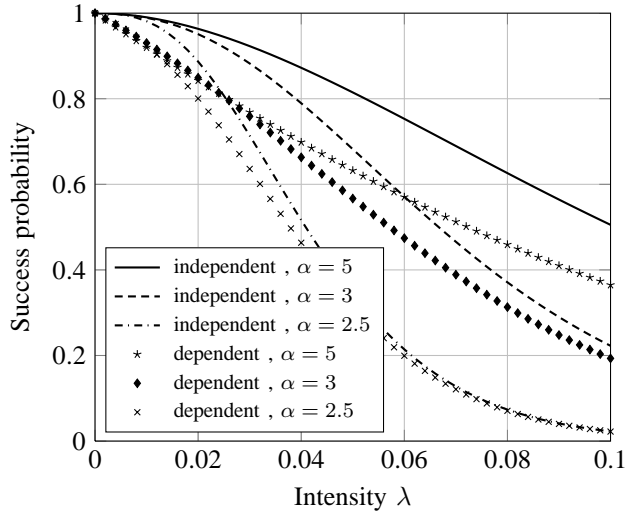

Finally, we investigate the probability that at least one out of two transmissions is successful, again for both the dependent and the independent interference case.
This scenario is sometimes denoted as time diversity or retransmission scenario.
For the dependent interference case, the probability of at least one successful transmission is given by $\prob[\sevent_\indiv]+\prob[\sevent_\indv]-\prob[\sevent_\indiv,\sevent_\indv]$. For the independent interference case, we can calculate this probability by $1-(1-\prob[\sevent_\indiii])^2$.
A plot of these probabilities is presented in Fig.~\ref{fig:Psucc:Nakagami:timediv}.
As can be seen, the success probabilities for independent interference are higher than the ones for the dependent interference. Note that it is the other way around in Fig.~\ref{fig:Psucc:Nakagami:dens:comp}. This can be explained by the fact that for highly correlated interference (which is the case in the plot due to $\txp=1$) it is very likely that either both transmissions are successful or both are not. Hence, the probability of having exactly one of two transmissions being successful is relatively low in this case. Therefore, the probability of at least one transmission being successful is also reduced.

Note that the success probabilities monotonically depend on $\thr$ and hence similar trends will occur for different values of $\thr$.
%Note that all observations made strongly depend on all other parameters. Hence, the effects observed with other parameter sets can be different.

\section{Concluding Remarks and Future Work}\label{sec:conclusion}

Interference is considered to be one of the key factors limiting the performance of distributed wireless networks. A good model of the interference and its space-time dynamics is an important asset for performance assessments. This paper contributes to this aspect in multiple ways. 

Firstly, we extended the toolbox of stochastic geometry to allow the calculation of very general functionals of PPPs. In particular, we proved a theorem that provides an expression for the general functional $\expect_{\ppp,\rndvar}\left[\prod_{\indi=1}^\fcount\left(\sum_{\pointi\in\ppp}\f_\indi(\pointi,\rndvar_\pointi)\right)^{\fexp_\indi}\prod_{\pointi\in\ppp}\g(\pointi,\rndvar_\pointi)\right]$. This result can be seen as an extension of the well-known Campbell-Mecke theorem for the PPP.

Secondly, we applied this general result, which has a broad range of applications, to interference in wireless networks. This allowed us to calculate the expected value $\expect\left[\prod_{\indi=1}^\fcount\interf_\indi^{\fexp_\indi}\,\exp\big(-\const \interf_\indi\big)\right]$. This result can be applied to different scenarios: Similar expressions occur, e.g., when calculating the outage probability of cooperative communications.

Thirdly, we highlighted one of these examples, namely calculating the joint outage probability of several transmissions under Nakagami fading. This derivation holds for any path loss model; as a case study, we presented the result for the singular path loss model. The intention is to sketch the path going from the general result to the final expression for a particular path loss model.

Our goal for future work is to further extend the tools of stochastic geometry for use in wireless settings. In particular, we will consider more sophisticated medium access, departing from ALOHA and aiming for CSMA. Modeling a CSMA network will involve hard-core point processes, for which fewer mathematical tools are available.

\section*{Acknowledgments}
This work has been supported by the Austrian Science Fund (FWF) grant P24480-N15 (Dynamics of Interference in Wireless Networks) and KWF/EFRE grant 20214/20777/31602 (Research Days), the European social fund, a research leave grant from the University of Klagenfurt, and Greek national funds through the program ``Education and Lifelong Learning'' of the NSRF program ``THALES investing in knowledge society through the European Social Fund''. Also the support of the U.S.~NSF (grant CCF 1216407) is gratefully acknowledged.

\bibliographystyle{ieeetr}    %in order of reference
%\bibliographystyle{unsrt}    %in order of reference
%\bibliography{bettstetter,cooperation,paper}
\bibliography{IEEEabrv,paper}

\appendices
\section{}\label{sec:app:lemmata}
\begin{lemma}\label{lem:exchange}
Let $f_\indi:\pppspace\to\re$ with $1\leq \indi\leq k$ denote non-negative functions and $\someset\subseteq\pppspace$ be a finite or countable set.
%\comment{Elements in $U$ are distinct.}
Then we have
\begin{equation}
\sum_{u\in \someset^k}\prod_{\indi=1}^kf_i(u_i)=\prod_{\indi=1}^k\sum_{u\in \someset}f_i(u)\:.
\end{equation}
\end{lemma}
\begin{IEEEproof}
The result holds due to the law of distributivity.
We prove the lemma by induction. For $k=1$ the result is trivial. Let us assume the result holds for $k$; we show that it also holds for $k+1$.
Thus, we have
\begin{eqnarray}
\prod_{\indi=1}^{k+1}\sum_{u\in\someset}f_\indi(u)&=&\left(\prod_{\indi=1}^{k}\sum_{u\in\someset}f_\indi(u)\right)\sum_{v\in\someset}f_{k+1}(v)\\\nonumber
&\stackrel{(a)}=&\left(\sum_{u\in\someset^k}\prod_{\indi=1}^{k}f_\indi(u_\indi)\right)\sum_{v\in\someset}f_{k+1}(v)\\\nonumber
&=&\sum_{v\in\someset}f_{k+1}(v)\left(\sum_{u\in\someset^k}\prod_{\indi=1}^{k}f_\indi(u_\indi)\right)\\\nonumber
&=&\sum_{v\in\someset}\sum_{u\in\someset^k}\left(f_{k+1}(v)\prod_{\indi=1}^{k}f_\indi(u_\indi)\right)\\\nonumber
&=&\sum_{u\in U^{k+1}}\prod_{\indi=1}^{k+1}f_\indi(u_\indi)\:,
\end{eqnarray}
where $(a)$ holds due to the induction assumption.
\end{IEEEproof}

%\input{old_results.tex}

% Bibliography

\end{document}